\documentclass{aa}
\usepackage{amsmath, amssymb}
\usepackage{natbib}
\usepackage{graphicx}
\usepackage{txfonts}
\begin{document}
   \title{About the relative importance of compressional heating and current dissipation for
   the formation of coronal X-ray Bright Points}
%{The role of adiabatic compression and Joule heating in the formation of X-ray Bright Points}

   \author{S. Javadi
          \inst{1}\fnmsep\thanks{\email{javadi@mps.mpg.de}}
           J. B\"uchner
           \inst{1}
           A. Otto
           \inst{2}
            \and
           J.C. Santos\inst{1}
          }
\offprints{S.Javadi}

   \institute{Max--Planck--Institut f\"ur Sonnensystemforschung, 37191 Katlenburg-Lindau, Germany
         \and
             Geophysical Institute, University of Alaska, Fairbanks, AK 99775, USA\\
             \email{ao@how.gi.alaska.edu}
%          \and
%             Instituto Nacional de Pesquisas Espaciais-INPE, AV. dos
%             Astronautas, 1758, Jd. da Granja, 12227-010, S\~{a}o Jos\'{e} dos Campos, S\~{a}o Paolo, Brazil\\
             }

%   \date{Received September 15, 1996; accepted March 16, 1997}

% \abstract{}{}{}{}{}
% 5 {} token are mandatory

\abstract
   { The solar corona is heated to high temperatures of the order of $10^{6} K $.
   The coronal energy budget and specifically possible mechanisms of coronal heating (wave, DC-electric fields, ..)
   are poorly understood. This is particularly true as far as the formation of X-ray
   bright points (BPs) is concerned. }
   { Investigation of the energy budget with emphasis on the relative role and contribution of adiabatic
   compression versus current dissipation to the formation of coronal BPs. }
   { Three-dimensional resistive MHD simulation % (jean: starting with a potential field extrapolated )
   starts with the extrapolation of the observed magnetic field from SOHO/MDI
   magnetograms, which are associated with a BP observed on 19
   December 2006 by Hinode. The initial radially non-uniform plasma density
   and temperature distribution is in accordance with an equilibrium
   model of chromosphere and corona. The plasma motion is included in the model as
   a source of energy for coronal heating.}
   { Investigation of the energy conversion due to Lorentz force, pressure gradient
   force and Ohmic current dissipation for this bright point shows the minor effect
   of Joule heating in comparison to the work done by pressure gradient force
   in increasing the thermal energy by adiabatic compression.
   Especially at the time when the temperature enhancement
   above the bright point starts to form, compressional effects are quite
   dominant over the direct Joule heating.   }
   { Choosing non-realistic high resistivity in compressible MHD models for simulation of solar corona
   can lead to unphysical consequences for the energy balance analysis, especially
   when local thermal energy enhancements are being considered.  }

\titlerunning{Compressional heating vs local dissipation in a X-ray Bright Point}
\authorrunning {S. Javadi et al. }

   \maketitle

%________________________________________________________________

\section{Introduction}
\label{introduction}

The mechanisms of coronal heating are not well understood.
A particular object for studying heating processes are coronal
bright points (further abbreviated BPs). Due to the increasing
accuracy of observations our knowledge about BP has greatly
advanced from the time of their discovery in soft X-ray images
\citet{vaiana}. According to X-ray and EUV observations the linear
size of BPs is on average about 30-40 arcsec with, typically, an
embedded bright core of about 5-10 arcsec  \citet{madjarska}. The
average lifetime of X-ray BPs is about 8 hours \citet{golub} and 20
hours for EUV BPs \citet{zhang}.
For a long time it has been known that BPs are associated with small
bipolar magnetic features in the photosphere
\citet[]{krieger, brown}. About one third of BPs lie over
emerging regions of magnetic flux, while the rest of them lie above
moving magnetic features. This was a base for the "cancelling
magnetic feature" (CMF) model \citet{priest94}. Lifetime and energy
release of BPs are known to be closely related to the different
phases of the motion of this photospheric magnetic feature
\citet{brown}.
% \citep[e.g.,][]{parnell94,longcope98}
First theories were mainly addressing the topology of the magnetic
field below BPs by e.g., \citet[]{parnell94,longcope98}. Using higher
resolution and cadence observations of BP's intensity and taking into account a
more comprehensive patterns of motion in particular in regions with
highly divergent magnetic field, \citep{brown} could associate
different patterns of motion of the solar photospheric magnetic
features to different stages of a BP evolution.
% \citep{Rekowski2006a,Rekowski2006b,Rekowski2008}
The plasma motion in the regions of strong magnetic field was first
included by B\"uchner (2004a,b) in their three-dimensional numerical
resistive MHD model using their 3D numerical simulation model,
LINMOD3d. The latter considers dissipation of currents generated by
plasma motion in photosphere on time scales longer than an Alfv\'en
time as a one of the heating processes in the solar corona
\citet{parker72}. In their model they took into account current
dissipation due to anomalous resistivity
\citep[]{BuechnerElkina:2005,BuechnerElkina:2006} that causes Joule
heating. Since LINMOD3d considers the compressibility of the plasma,
the resulting heating could be due also to compressional effects.
Later on two-dimensional MHD simulation studies were carried out by
\citet{Rekowski2006a,Rekowski2006b,Rekowski2008}. These authors used
an analytical initial equilibrium and imposed a magnetic flux
footpoint motion to model coronal bright point heating as being due
to canceling magnetic features. To obtain the desired heating rate
they used an enhanced resistivity for which the values were above
the theoretically justifiable resistivity. This raises the general
question of the energy budget and energy conversion in solar flux
tubes. Even with low resistivity, current simulations are unable to
resolve the diffusion regions of reconnection and thus overestimate
Joule heating. It is also unresolved how much heating is caused by
pressure gradient forces.

To clarify this question we continued the work of
\citet{buechner04a,buechner04b,buechner04c,buechner06,buechner07,santos07,santos08}.
These authors demonstrated the formation of localized current sheets
in and above the transition region at the position of a EUV BPs as a
result of photospheric plasma motion. This study is extending their
results through a systematic study of the energy conversion and
budget in magnetic flux tubes. The investigation uses the 3D
simulation model LINMOD3d to simulate the solar atmosphere in the
region of an X-ray BP observed by the Hinode spacecraft on 19
December 2006 between 22.17 UT and 22.22 UT.

In section~\ref{model} we briefly review the main features of the
numerical simulation model LINMOD3d. In section~\ref{setup} we
describe the specific simulation setup used in our study and
section~\ref{results} provides some simulation results for the
chosen BP data. In section~\ref{energy} we present results of energy
budget analysis by investigating the role of different forces and in
section~\ref{summary} we summarize and discuss our results.
%__________________________________________________________________

\section{Simulation model}
\label{model} Our simulation model uses the approach of the LINMOD3d
code \citep[]{buechner04a,buechner04b,buechner04c}. This means that
the initial magnetic field is obtained by extrapolating the observed
photospheric line-of sight (LOS) magnetic fields. The initial plasma
distribution is non-uniform containing a dense and cool chromosphere
as well as the transition to a rarefied and hot corona.
%(Before starting the actual simulation the configuration is relaxed to a force-free equilibrium. After that t)
The photospheric driving is switched on by coupling
the chromospheric plasma with a moving background neutral gas. Some details of our code have been
given briefly in the following subsection.

\subsection{Equations}

In our study we solve the following set of MHD equations:
\begin{eqnarray}
        \frac{\partial \rho}{\partial t} &= - \vec \nabla \cdot \rho \vec u \\
        \frac{\partial \rho \vec u}{\partial t} &= -\vec \nabla \cdot \rho \vec u \vec u - \vec \nabla p + \vec j \times \vec B - \nu \rho
             (\vec u - \vec u_{0}) \\
        \frac{\partial \vec B}{\partial t} &= \vec \nabla \times (\vec u
             \times \vec B - \eta \vec j)\\
        \frac{\partial p}{\partial t} &= - \vec \nabla \cdot p \vec u -
            (\gamma -1) p \vec \nabla \cdot \vec u + (\gamma -1) \eta j^{2}
\end{eqnarray}

where $\rho$ and $\vec u$ are plasma density and velocity, $\vec B$
is the magnetic field and P is the thermal pressure. A
plasma-neutral gas coupling in photosphere and chromosphere is
included through the collision term in the momentum equation, where
$\vec u_{0}$ denotes the neutral gas velocity. The neutral
gas serves as a frictional background to communication photospheric
footpoint motion to the plasma and magnetic field through frictional
interaction. It also leads to a reflection of coronal Alfv\'en waves
back to the corona from the transition region, so that the influence
of coronal Alfv\'en waves can be neglected at the photospheric
boundary. In order to set the plasma in motion a number of
incompressible flow eddies is used according to observed horizontal
drifts in the photosphere $ \nabla \cdot \textbf{u}_{0} = 0$
is imposed via the neutral gas, where $\textbf{u}_{0}$
is dependent in x and y. It is constant along z and derived
from a potential using $\textbf{u}_{0} = \nabla \times (U
\textbf{e}_{z})$, with
  \begin{equation}
 U = u_{00} / \cosh\left(\frac{x - y + c_{0}}{L_{0}}\right) / \cosh\left(\frac{x + y + c_{1}}{L_{1}}\right)
 \end{equation}

Note that the contour lines of this function are streamlines
of the flow. The magnitudes of velocity scale with $u_{00}/L_{0}$
and $u_{00}/L_{1}$, chosen in accordance with the observed plasma
motion in the photosphere. In our simulation we approximated the
observed motion by three vortices with amplitudes of the velocity
$u_{00}$ equal to 5.5, 5 and 2 $km/s$, respectively. The values of
$c_{0}$, $L_{0}$, $c_{1}$ and $L_{1}$ are 9, 6, 51 and 6 Mm for the
first vortex 5, 6, 28 and 6 Mm for the second and 19, 7, 38 and 7 Mm
for the third vortex. The height-dependent collision
frequency $\nu$ is chosen to be sufficiently large only below the
transition region. This way the plasma is forced to move, dragged by
the neutral gas, in the model chromosphere but not above the
transition region. This way the horizontal motion generates a
Poynting flux into the corona. On the other hand the collision
frequency is chosen in a way that coronal Alfv\'en waves are
properly reflected while wave perturbations in the chromosphere are
heavily damped by the frictional interaction with the neutral
background. Our choice of equations means that in this study we do
not consider energy losses due to radiation and heat conduction and
we also excluded the action of the solar gravitation in this study.
The system of equations is closed by Ohm's and Amp\`ere's laws and
the temperature is defined via the ideal gas law for a fully ionized
plasma:

\begin{eqnarray}
         \vec E                       &= - \vec u \times \vec B + \eta \vec j  \\
         \vec \nabla \times \vec B    &= \mu_{0} \vec j      \\
         p                            &= 2n\kappa_{B}T
\end{eqnarray}

The value of the resistivity $\eta $ is varied in accordance with
three models described in subsection \ref{resistivitymodels}. The
MHD equations are discretized by means of a  second order weakly
dissipative Leapfrog scheme. Due to stability reasons the induction
equation is discretized using Dufort-Frankel scheme, \citet{potter}.

\subsection{Simulation box and normalization}

The lower boundary of the simulation box is a horizontal square in
the photosphere sized $46.4 \times 46.4 \,  Mm^{2}$. The simulation
box extends 15.45 Mm toward the corona. A nonuniform grid in the z
direction supplies the proper resolution of the transition layer,
where the grid distance $\Delta z $ corresponds to 160 km,
\citet{buechner04a}. This corresponds to 64 grid points in z
direction, while in the x, y plane a $128 \times 128$ grid are used.
We solve for dimensionless variables that are normalized to natural
scales as listed in table.~\ref{norm}.
Note that the maximum imposed velocity of the neutral gas is smaller than 5 km/s while the
typical (normalizing) electron thermal velocity is $v_{the} = 1470$
km/s and the Alfv\'en speed is $v_{A} = 50$ km/s. Hence, one can be
certain that the inserted neutral gas motion is gentle,
sub-Alfv\'enic and sub-slow  velocities.

\subsection{Resistivity models}
\label{resistivitymodels} In order to verify the influence of
different resistivity models on the BP plasma heating we solved the
equations for the same initial and boundary conditions but varying
the resistivity model. The resistivity $\eta$ can be
expressed via an effective collision frequency $\mu$ as $\eta =
\frac{\mu}{ \epsilon_{0} \ \omega^{2}_{pe}}$, where $\omega_{pe}$ is
the electron plasma frequency ($\omega_{pe} = \sqrt{n e^{2}/
\epsilon_{0} m_{e}}$). In our model we always apply a constant
physically justified background resistivity $\eta_{0}$ which exceeds
exceeds the numerical resistivity. It is appropriate to chose for
effective collision  frequency of the background resistivity the
\citet{spitzer} value $\mu = (n e^{4} Ln \Lambda T^{-\frac{3}{2}} /
16 \pi \epsilon^{2}_{0} m^{\frac{1}{2}}_{e} K^{\frac{3}{2}}_{B}$).
Based on the typical plasma parameters of our model we chose for the
collision-driven background resistivity $\eta_{0} = 10^{-4}$ (in
normalized units). In two models we switched on additional,
anomalous, resistivity in places where either the current density of
the current carrier velocity ($u_{ccv}$ determined as the current
density divided by the charge density) exceeds a physically
justified thresholds of micro-instabilities.
\\
In the first resistivity model anomalous resistivity is switched on
when the current carrier velocity ($u_{ccv}$ exceeds a critical
velocity \citep[]{roussev02,BuechnerElkina:2005}:

\begin{eqnarray}
\eta = \eta_{0} +
       \begin{cases}
                0,  & \text{if} \, |u_{ccv}| < u_{crit}  \\
            \eta_{eff} \left( { \frac{|u_{ccv}|}{u_{crit} } - 1 } \right),
            & \text{if} \, |u_{ccv}| \geq u_{crit} \\
       \end{cases}
\end{eqnarray}
% bold
A natural choice for the threshold velocity is the electron
thermal velocity $v_{the}$, in our for the normalizing quantities
1470 km/s or to $5.8. 10^{-4}$ in normalized units. In the first
resistivity model we chose $5.10^{-2}$ to follow the ideal evolution
of the plasma as long as possible. The additional term for
resistivity can be estimated e.g., for a nonlinear ion-
acoustic instability \citep{BuechnerElkina:2006} as

\begin{equation}
\eta_{eff} =  \frac{\mu_{eff}}{ \epsilon_{0} \ \omega^{2}_{pe}} = \frac{\omega_{pi}}{\epsilon_{0} \ \omega^{2}_{pe}}
\end{equation}
% bold
Here $\omega_{pi}$ denotes plasma ion frequency ($\omega_{pi} = \sqrt{n e^{2}/ \epsilon_{0} m_{i}}$). For
the typical parameters of our simulation this estimate would reveal
$\eta = 2.5$, i.e. a magnetic Reynolds number of less than unity. In
this case many current sheets would immediately diffuse away. On the
the other hand, since the plasma $\beta$ is relatively large for our
simulation parameters obliquely propagating waves would be present
in the spectrum of the micro-turbulence. In this case the estimate
of the effective collision frequency has to take into account
lower-hybrid waves \citep{silin}. For our normalizing values this
results in $\eta_{eff}= 0.03$.

In a second model calculation we considered a current density
dependent resistivity used before, e.g., by \cite{neukirch}, in which the resistivity increases even stronger
(quadratic dependence) after the current density exceeds a critical
value $j_{crit}$:

\begin{eqnarray}
\eta = \eta_{0} +
   \begin{cases}
      0, & if |j| < j_{crit}  \cr
      \eta_{eff}( \frac{|j|}{j_{crit}} - 1)^2 & if |j| \geq j_{crit} \cr
   \end{cases}
\end{eqnarray}
The critical current density is related to critical velocity
via $j_{crit} = e \ n_{e} \ u_{crit}$. Here we will report the
results of our simulations obtained according to the second model
for which we chose a threshold as low as $j_{crit} = 0.69$ in order
to discuss the consequences of an early switch on of additional,
anomalous resistivity. For comparison we solved the problem also by
assuming for a third model a constant enhanced uniform resistivity
as usually done in global MHD simulations.

Concerning the values of the chosen $\eta_{eff}$ one should
note that the width of the actual current sheets in which turbulence
effectively operates is of the order of the ion inertial scale $d_i
= {c \over \omega_{pi}}$. This scale cannot be resolved in any
realistic 3D MHD simulation of the solar corona. In order to
introduce micro-turbulent anomalous resistivity the threshold
velocity (- current density) has to up-scaled to the actual
resolution of the simulation by a factor of $5. 10^{4}$. By the same
reason the resistive electric field builds up in very (perhaps
$d_i$-) thin current sheets. To consider the correct values of the
electric field on the much coarser MHD-simulation grid anomalous
resistivity used in the simulation has also to be scaled up by the
above scaling factor. This approach allows to consider the correct
amount of Joule heating.
%\begin{equation}
%\eta = \eta_{0} +  10 \  \eta_{0} \ (1 - exp(- \frac{ time}{ 0.5}) )
%\end{equation}
%
%__________________________________________________ One column table
   \begin{table}
      \caption[]{ Normalization values.}
         \label{norm}
     $$
         \begin{array}{p{0.5\linewidth}l}
            \hline
            \noalign{\smallskip}
            Variable      &  \ Normalization\ value \\
            \noalign{\smallskip}
            \hline
            \noalign{\smallskip}
%           Yorke 1979, Yorke 1980a & \leq 1700             \\
%            Chosen values:                              \\
\\
            density    &    N_{0} = 2. 10^{15}  m^{-3}  \\
\\
            lenght     &    L_{0} = 500 \ km     \\
\\
            magnetic field   &   B_{0} = 1 G = 10^{-4}\ T     \\
             \ldots\ldots\ldots\ldots\ldots\ldots\ldots\ldots\ldots\ldots\ldots\ldots\ldots\ldots\ldots\ldots\ldots\ldots\ldots\ldots\ldots\\
%             Calculated \ values:                \\
            pressure       &    P_{0} = \frac{B_{0}^{2}}{2 \mu_{0}} = 4 \times 10^{-3} \ J/m^{2}      \\
\\
            temperature    &    T_{0} = \frac{P_{0}}{2 \ n_{0} \ \kappa_{B}} = 7.2 \times 10^{4}\ K     \\
\\
            Alfv\'en velocity   &    v_{A0} = \frac{B_{0}}{\sqrt{\mu_{0} \ m_{i}\ N_{0}}} = 50 \ km/s     \\
\\
            time    &    \tau_{0} = L_{0} / \ v_{A0} = 10 \ s    \\
% \\
%            thermal velocity  &  v_{th0} = v_{A0} / \sqrt{2} = 35.36 \ km/s     \\
%             \ldots\ldots\ldots\ldots\ldots\ldots\ldots\ldots\ldots\ldots\ldots\ldots\ldots\ldots\ldots\ldots\ldots\ldots\ldots\ldots\\
%            Tranfport\ coeffiecients:              \\
%\\
%             $\hat{\chi} = \frac{2 \ n_{0} \ L_{0}}{m_{p} \ v_{A_{0}}^{3}} \ \chi$  &  \rightarrow \chi_{0} = (\frac{m_{p} \ v_{A_{0}}^{3}}{2 \ n_{0} \ L_{0}}) / k^{\alpha} = \ 10^{-34}/ k^{\alpha} \\
%\\
%             $\hat{\kappa}_{\parallel}=\frac{10^{-11} \ T_{0}^{5/2}}{2 \ v_{A0} \ n_{0} \ \kappa_{B} \ L_{0}} \ \hat{T}^{5/2}$ & \rightarrow \kappa_{\| 0}=...? \\
%\\
%             $\hat{\kappa}_{T}=\tau_{A0} \ \kappa_{T}$ & \rightarrow \kappa_{T0} = \tau_{A0} \\
 %           \noalign{\smallskip}
            \hline
         \end{array}
     $$
   \end{table}

\subsection{Initial and boundary conditions}
\label{initial} We first carried out a potential field extrapolation
to the Fourier decomposed normal field component of the magnetic
field taken from the MDI-observation. The resulting 3D magnetic
configuration is used as the initial condition of the simulation
code. In the potential field approximation the normal field
component is related to $(B_{x}, B_{y})$ through $\nabla \cdot B =
0$ and $\nabla \times B = 0 $. The initial
density and temperature height profiles for the plasma is taken in
accordance with the VAL model that assumes pressure being in a
hydrostatic equilibrium. The simulation box has 6 boundaries: 4 lateral, 1
top and 1 bottom boundaries. For the side boundaries a line symmetric
boundary condition is used with the line symmetry with respect to the
centers of the sides of the simulation box. For the upper boundary
the derivatives in the normal direction are put to zero. At the lower
boundary the normal velocity is set to be zero, while the
tangential velocity is taken from the neutral motion.

\section{Simulation setup}
\label{setup} 
\begin{figure}[htp]
   \centering
   \includegraphics[width= 8.6 cm]{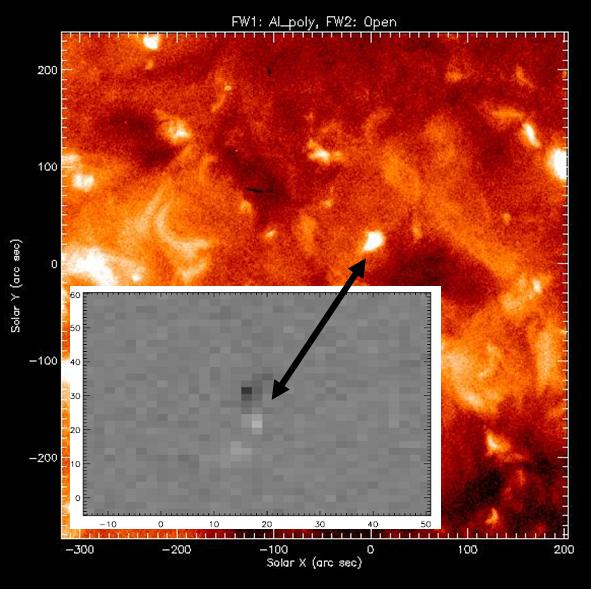}
   \caption{ X-ray Image taken from XRT/Hinode on 19 December 2006 at 22:08 UT.
  % The BP-region considered in this paper is indicated by a solid line square.
   Insert: the LOS component of the photospheric magnetic field in a 64 $\times$ 64 $arcsec^2$ horizontal
   plane taken from MDI/Soho, where white (black) spots correspond to upward (downward)
   directed Line-of-sight components of the photospheric magnetic field.
   The BP and the related magnetic field feature are indicated in the images. }
              \label{xrt_mdi4}%
    \end{figure}

Our study is based on an X-ray BP observed by the
XRT X-ray telescope on board of the Hinode spacecraft on 19 December
2006. The corresponding X-ray image is shown in Fig.~\ref{xrt_mdi4}.
For the initial magnetic field we used the observed  $line-of-sight$
(LOS) component of the photospheric magnetic field taken by the
Michelson Doppler Interferometer MDI onboard the Soho spacecraft at
22:17 UT. For that sake data from a field of view with the horizontal size
of 64 $\times$ 64 $arcsec^2$ was chosen that properly covers the magnetic
features associated to this BP (insert in Fig.~\ref{xrt_mdi4}). Note
that we use the LOS component as the initial normal field component
at the lower boundary of our simulation box, the photosphere, since
the BP observation was made close to the center of the solar disc.

Fourier filtering was applied to the LOS component of the magnetic
field. By taking into account only the first eight Fourier modes,
details of  magnetic field structure smaller that 6 Mm are
neglected. The extension of structures arising from smaller
scale magnetic features would not extend higher up into the corona,
they are dissipated at an early stage of the evolution in the highly
collisional chromospheric plasma.

\begin{figure}[htp]
   \centering
   \includegraphics[width= 8.6 cm]{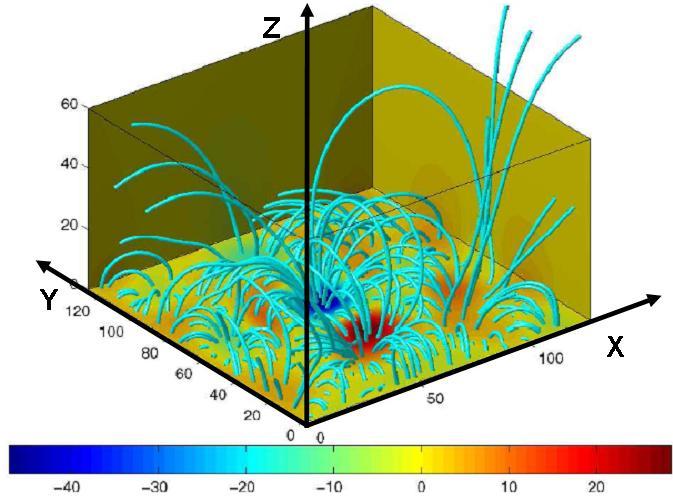}
   \caption{Potential magnetic field extrapolated from the filtered MDI
   magnetograms and used as initial configuration for the magnetic field in our simulation.
   The blue lines show the magnetic field lines. The color code depicts the LOS component of the magnetic
field. Note that axes here are in
   terms of grid points, 64 in z direction and $128 \times 128$ in x,y plane.}
     \label{extrapxyz}%
    \end{figure}

Fig.~\ref{extrapxyz} shows a three-dimensional view of the %(potential)
magnetic field extrapolated from the photospheric boundary for the magnetic field observed at 22:17 UT on
   December 19, 2006. The blue lines show the magnetic field lines. The color code depicts the LOS
   component of the photospheric magnetic field. Magnetic fields directed upward from the photosphere
   are colored in red, downward directed in blue.

\begin{figure}[htp]
   \centering
   \includegraphics[width= 8 cm, height = 0.26 \paperheight]{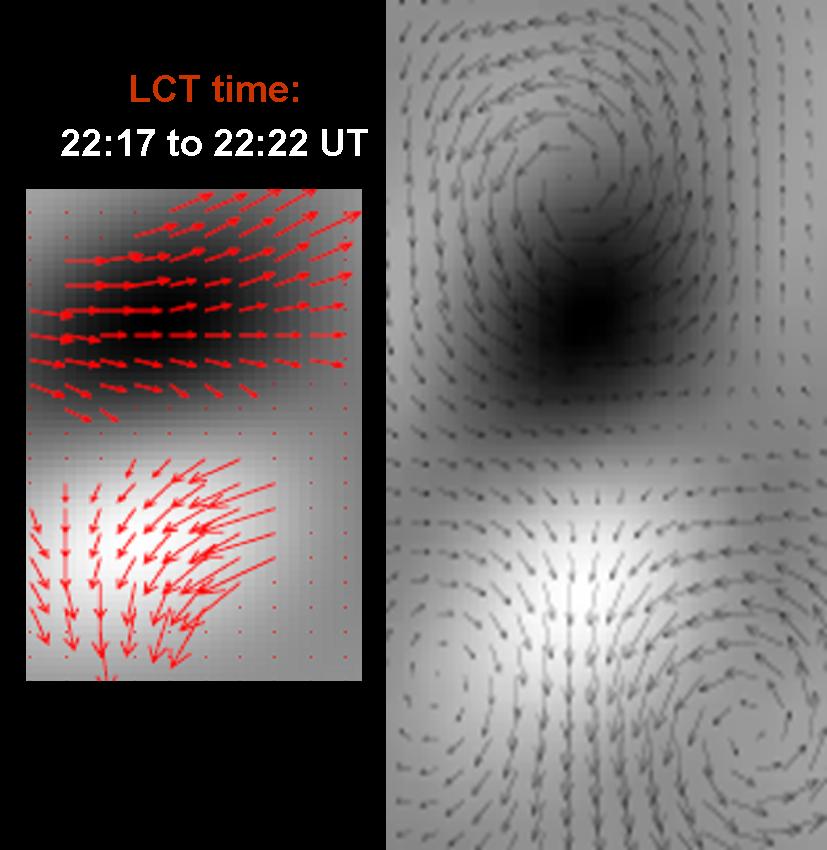}
   \caption{Horizontal plasma velocities in the photosphere. Left panel:
   velocities obtained by applying LCT technique to MDI magnetograms between 22:17 and 22:22 UT.
   The right panel shows the vortices that are used to approximate this motion in the simulation.   }
              \label{vortice5}%
    \end{figure}

With the chosen normalization length of $L_{0} = 500$ km, the box size in x
and y direction correspond to 92.8 $L_{0}$  and the z direction extend to 30.9 $L_{0}$.
The photospheric plasma velocities are obtained by applying the
local-correlation-tracking(LCT) method \citet{november88} to the
Fourier filtered LOS magnetic component of the
photospheric  magnetic field observed between 22:17 UT and 22:22 UT. The left panel of Fig.~\ref{vortice5}
shows the velocity pattern obtained by the LCT method. For the
simulation we used incompressible velocity vortices to mimic the observed velocity pattern, as shown in the
right panel of Fig.~\ref{vortice5}. Note that the interval chosen for the simulation starts a few hours
after the time the BP first appeared in the X-ray images and that
the bright point continues to glow a few more hours afterwards.
During the whole simulation time interval the relative shear motion
of the two main magnetic flux concentrations of opposite polarity is negligible.

\section{Simulation results}
\label{results} 
\begin{figure}[h]
          \centering
          \includegraphics[width=8.8 cm]{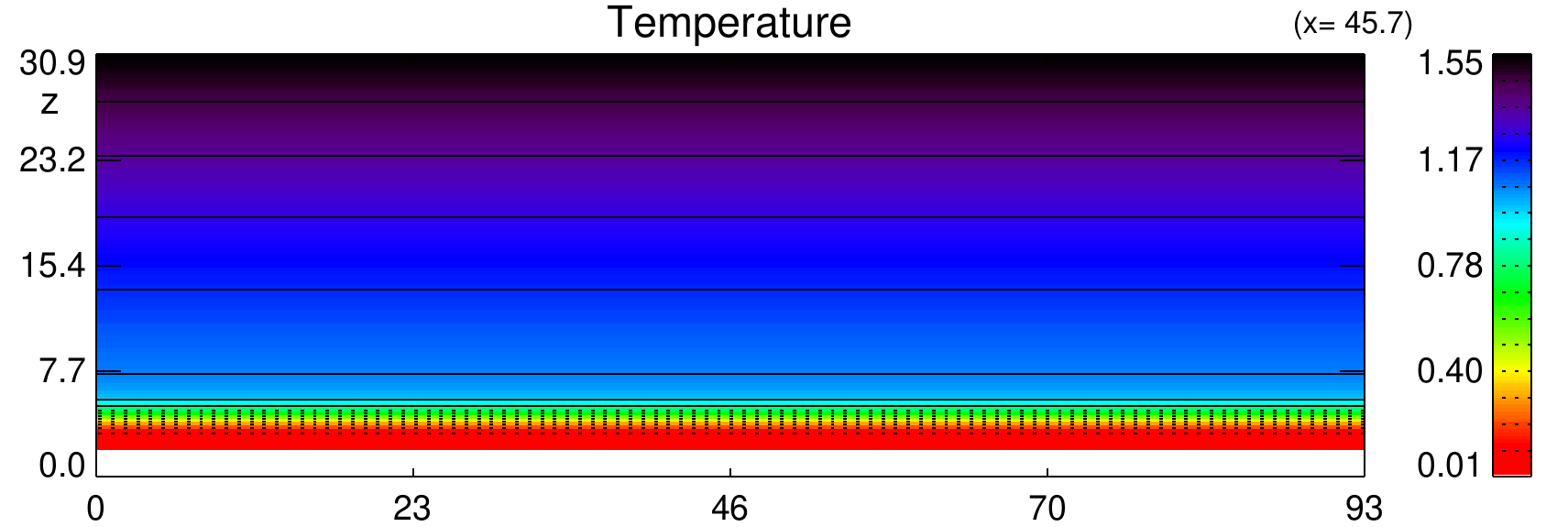}
 \includegraphics[width=8.8 cm]{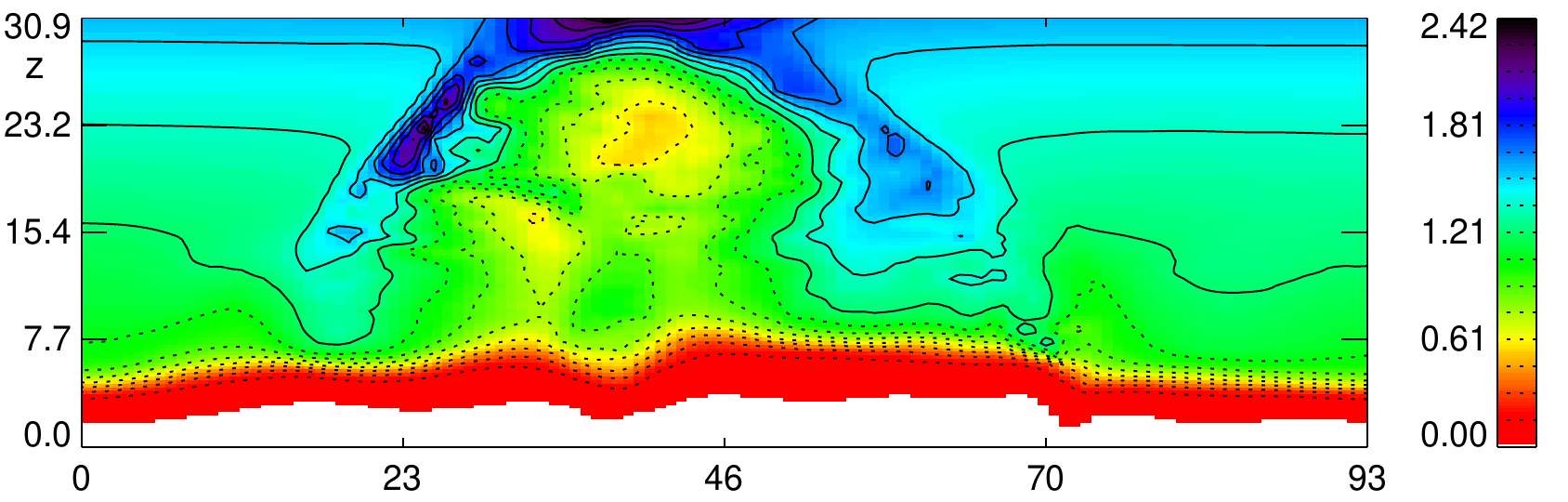}
  \includegraphics[width=8.8 cm]{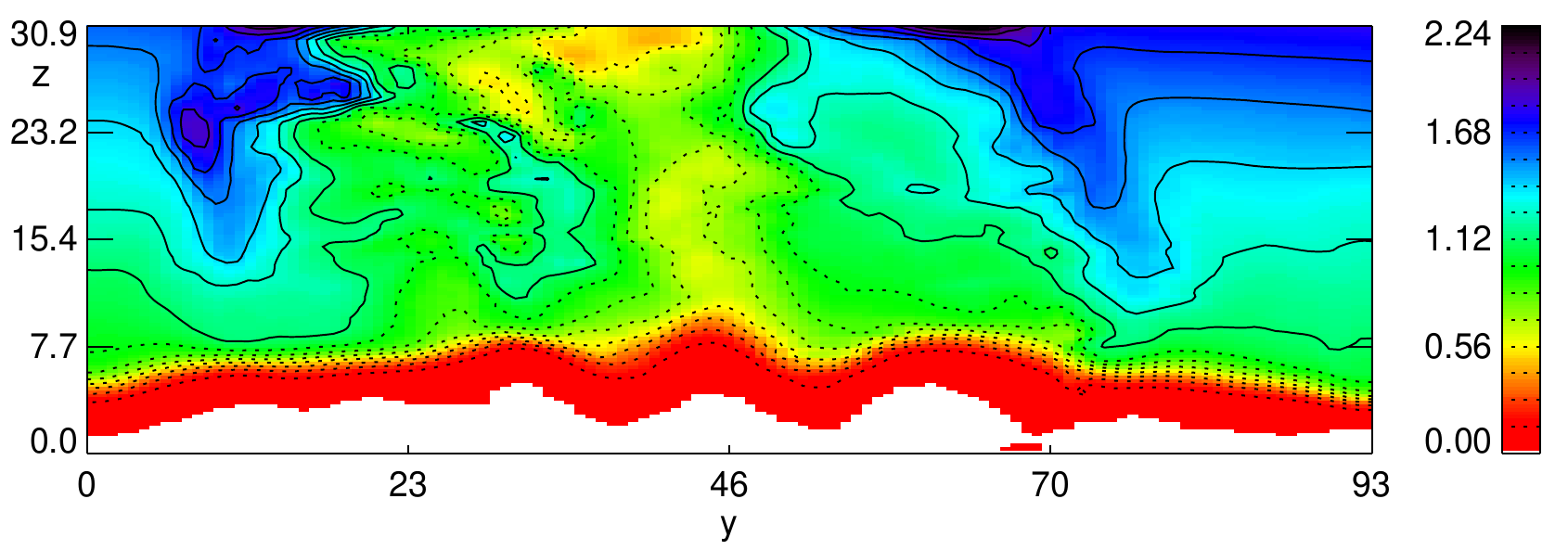}
          \caption{Temperature distribution in the vertical plane at x = 45.7 at
          the beginning (upper panel) and at t = 80 s  and t = 160 s.
           Note the temperature in color bar in presented in terms of the normalization value.  }
             \label{T_evol06816}%
         \end{figure}

The simulation results are first shown in a plane at
x = 45.7 (Fig.~\ref{extrapxyz}), which crossed through the center of
the two main magnetic polarities. The vertical profile of the
temperature is shown in Fig.~\ref{T_evol06816} for t = 0 (top
panel), 80 (middle panel) and 160 s (bottom panel). In $t = 0$ we
have a height dependent temperature as defined by the initial
condition. At $t = 80$ s the effects of plasma compression and
expansion, together with Joule heating, shape the temperature
profile. An arc of hot plasma is formed above the two opposite
magnetic polarities. The increase in temperature in this layer is
approximately 0.5 in normalized units, what corresponds to 36000 K.
The region that is located just below it, however, experiences some
drop in temperature. At $t = 160$ s the arc of hot plasma leaves the
simulation box and we are left with a corona in which the
differences in temperature can reach one orders of magnitudes.

\begin{figure}[h]
   \centering
   \includegraphics[width= 8.8 cm]{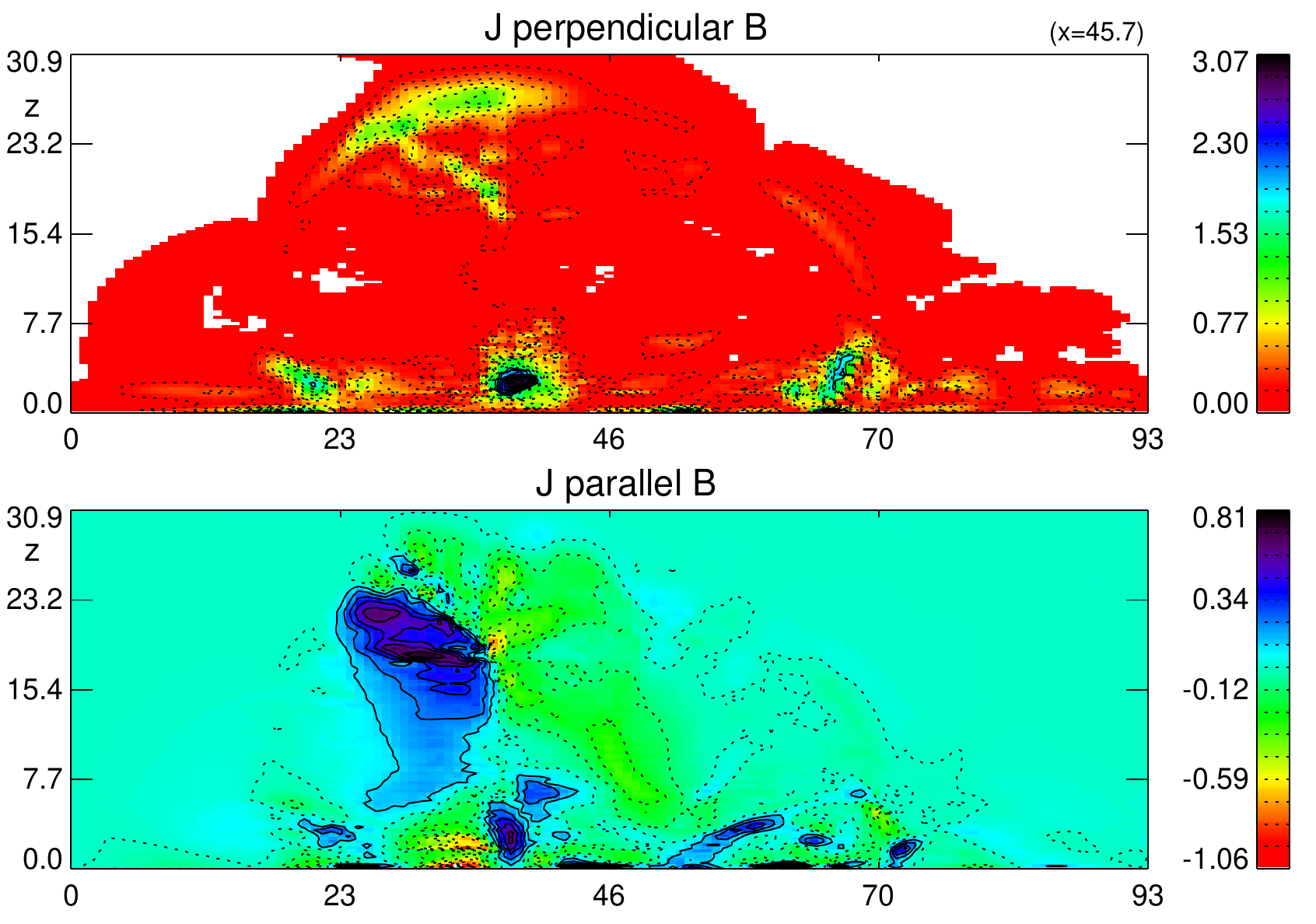}
   \includegraphics[width= 8.8 cm]{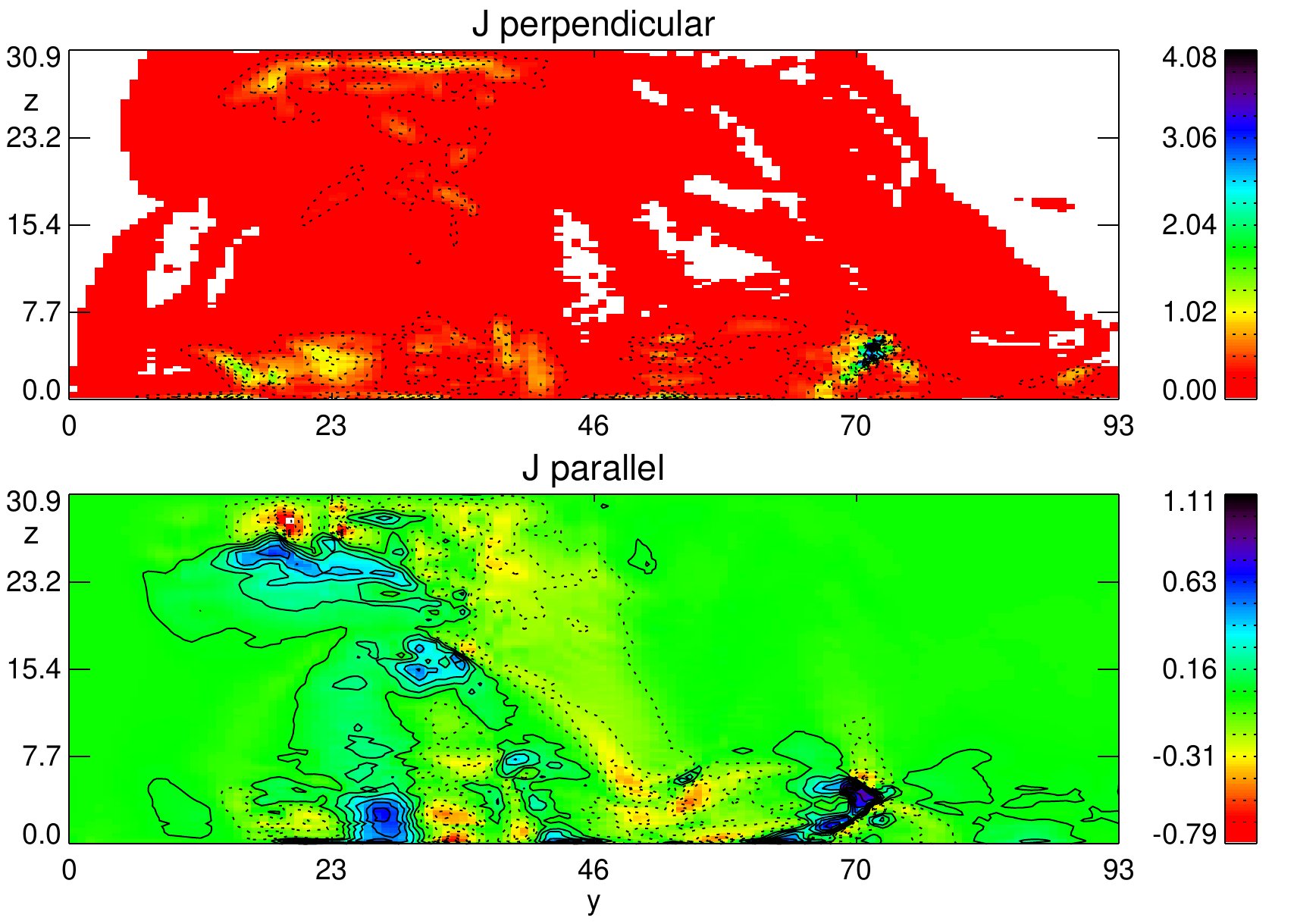}
   \caption{ Parallel and perpendicular components of electrical current at
   t = 80 s (upper two panels) and t = 160 s (lower two panels) in the same diagnostic plane as in Fig.4. Note the
   enhancement in perpendicular current is located at the same place of the
   temperature maximum.   }
              \label{Jpar_perp816}%
    \end{figure}

Fig.~\ref{Jpar_perp816} shows the parallel and perpendicular components of
the current with respect to the magnetic field direction at t = 80 and t =
160s. It can be seen that enhanced current flows coincide well with the
temperature increase. This would lead to an interpretation of the heating as
being due to current dissipation only. However, as shown later,
adiabatic heating can have an important contribution to temperature increase.

%_____________________________________________________________
% to have t and j in one fig, use T_J.jpg
%    \item time profile of the averaged currents \\ (i removed it)
%    The time profile of the averaged currents is shown in Fig.~\ref{Jprof}. The
%    enhancement of the perpendicular current in the heights below transition
%    region can be seen in the lower panel, that..
%
%    \begin{figure}[htp]
%    \centering
%    \includegraphics[width= 7 cm]{Jprof}
%    %%%\includegraphics{empty.eps}
%    \caption{ Time profile of the averaged J versus height... Enhancement of
%    the $J_{\perp}$ in the heights below transition region can be seen. }
%               \label{Jprof}%
%     \end{figure}
%
%    \end{itemize}
%_____________________________________________________________

 \section{Energy balance}
\label{energy}

Let us now diagnose the different contributions to plasma heating in
the BP region. First, in subsection~\ref{global}, we discuss the overall global
heating. In subsection~\ref{resistivity} the dependence on the resistivity model is presented. Finally, the
flux-tube heating is analyzed in subsection~\ref{fluxtube}.

 \subsection{Global effect of current dissipation and compression}
\label{global}

\begin{figure}[htp]
   \centering
   \includegraphics[width= 8.8 cm]{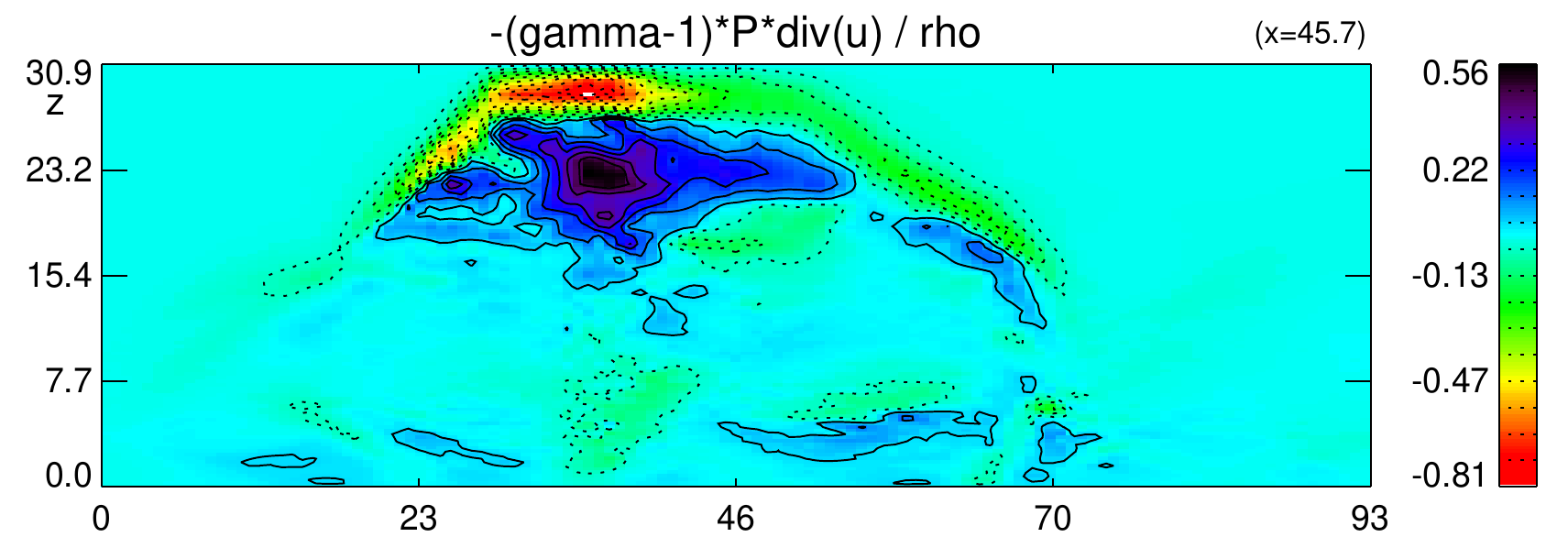}
    \includegraphics[width= 8.8 cm]{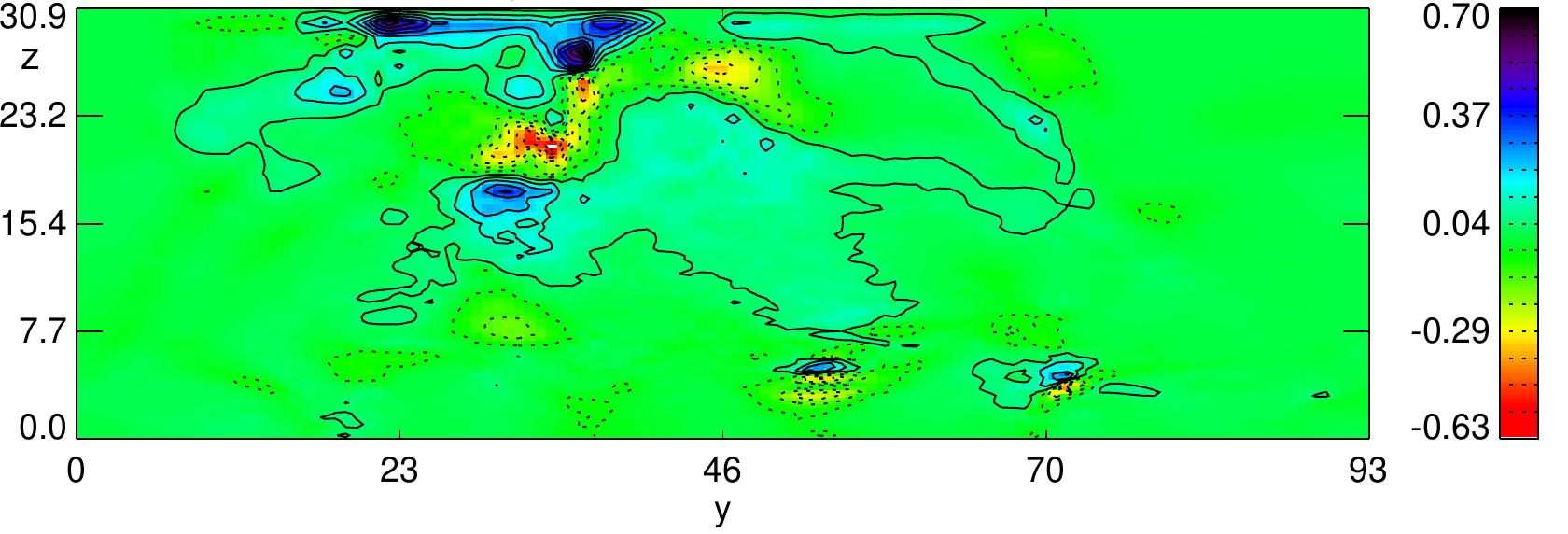}
   \caption{Adiabatic cooling/heating rate after 80 s and 160 s in the plane
    x = 45.7, according to the first term in the r.h.s of Eq. 11 over density, ($-(\gamma -1) T \nabla \cdot u < 0$).  }
    \label{Pexpand08816}%
%        \label{T_P}%
\end{figure}

In order to understand the relative contribution of current
dissipation and plasma compression to the coronal plasma heating in
the BP region it is appropriate to analyze the pressure changes rewriting the Eq. 4 in terms of a continuity
equation for the Temperature evolution. This leaves two source terms on
the right hand side of the equation:

\begin{equation}
%           \frac{\partial p}{\partial t} + u \cdot \nabla p = -\gamma p \nabla \cdot u + (\gamma - 1) \eta j^{2}
            \frac{\partial T}{\partial t} + \vec \nabla \cdot T \vec u = -(\gamma -1) \ T \ \vec \nabla \cdot \vec u + (\gamma -1) \ \eta j^{2} / \rho
\end{equation}
% \frac{P}{\rho}
Let us first analyze the first case, where an anomalous resistivity is used
when the current carrier velocity exceeds a critical value.
Fig.~\ref{Pexpand08816} shows the resulting distribution of $ -(\gamma -1)\ T
\ \vec \nabla \cdot \vec u $ in the vertical diagnostic plane. This way we
have a proxy for temperature changes associated to pressure compression and
expansion.

Adiabatic heating has an important role on the formation of the high
temperature arc that propagates upward towards the top boundary. It is also
due to expansion that temperature decreases below this hot arc.

\begin{figure}[htp]
   \centering
   \includegraphics[width= 8.8 cm]{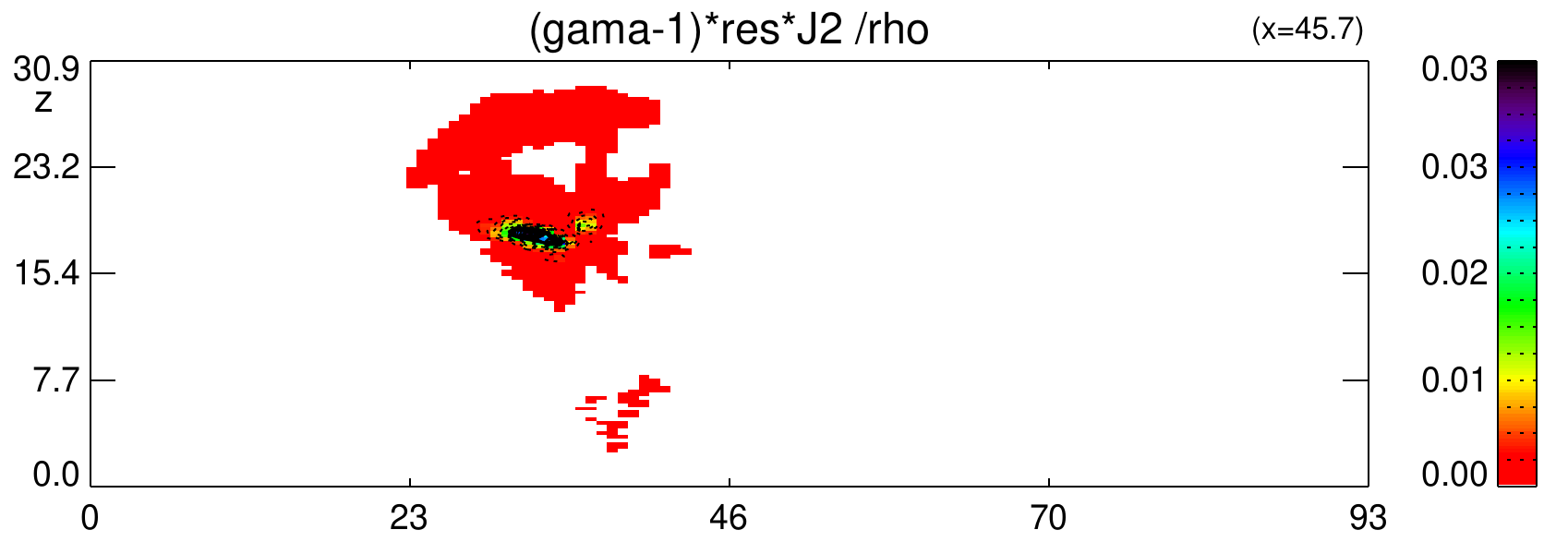}
      \includegraphics[width= 8.8 cm]{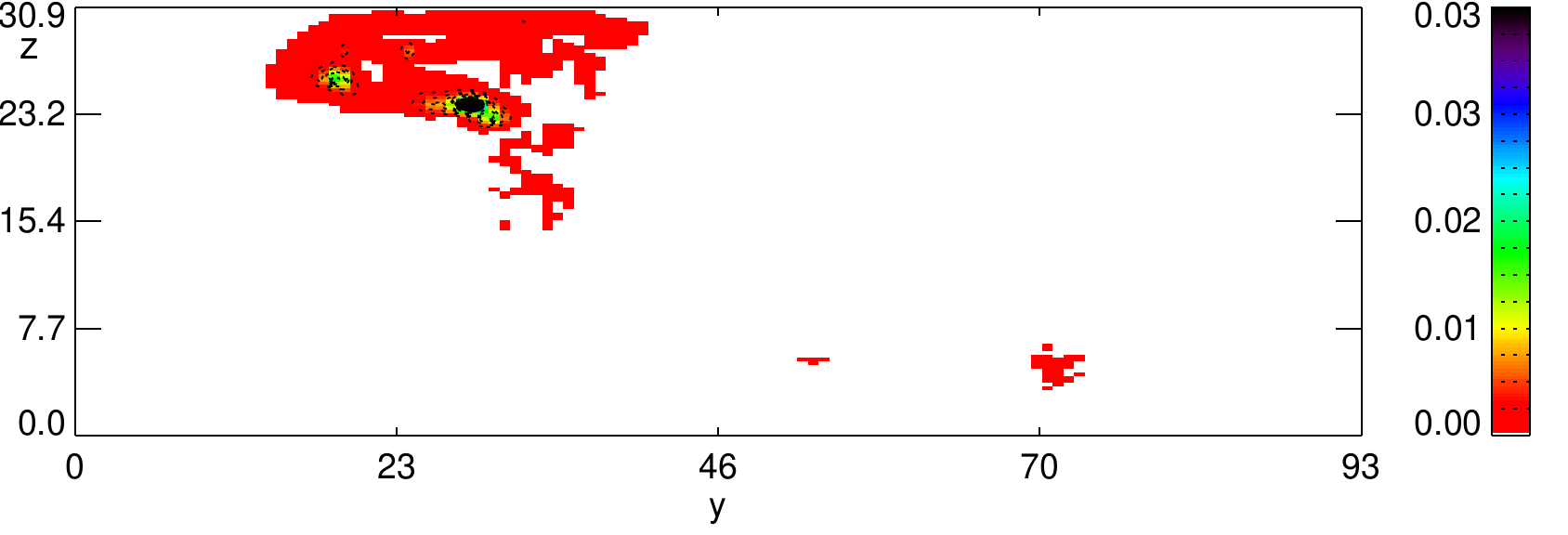}
   \caption{Joule heating rate, (second term in the
   r.h.s of Eq. 12, divided by density), after 80 s and 160 s in the plane x = 45.7.    }
   \label{gamresj2816}%
%   \label{T_J}%
\end{figure}

\begin{figure*}[htp]
   \centering
\includegraphics[width= 0.25 \paperwidth]{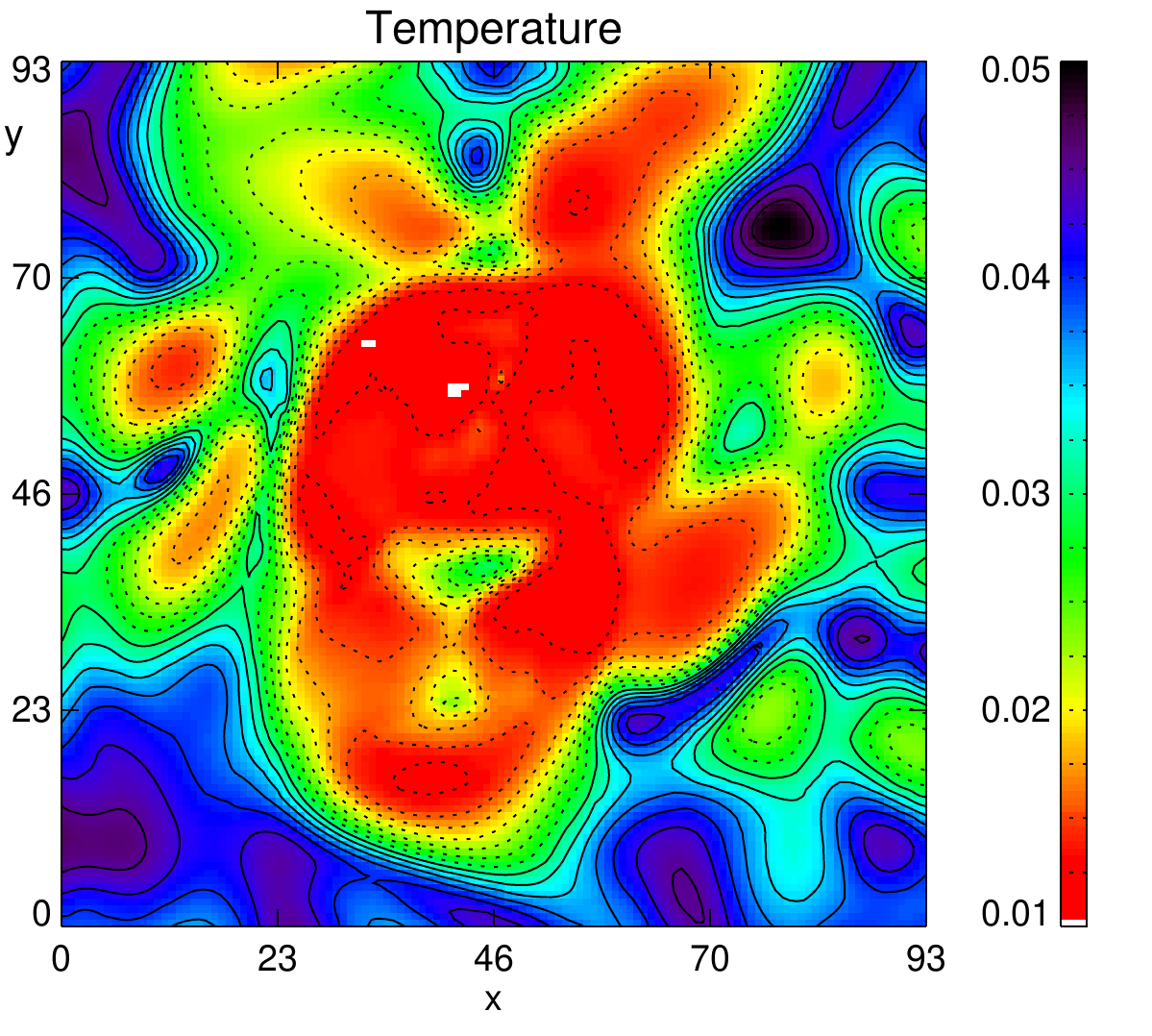}
\includegraphics[width= 0.25 \paperwidth]{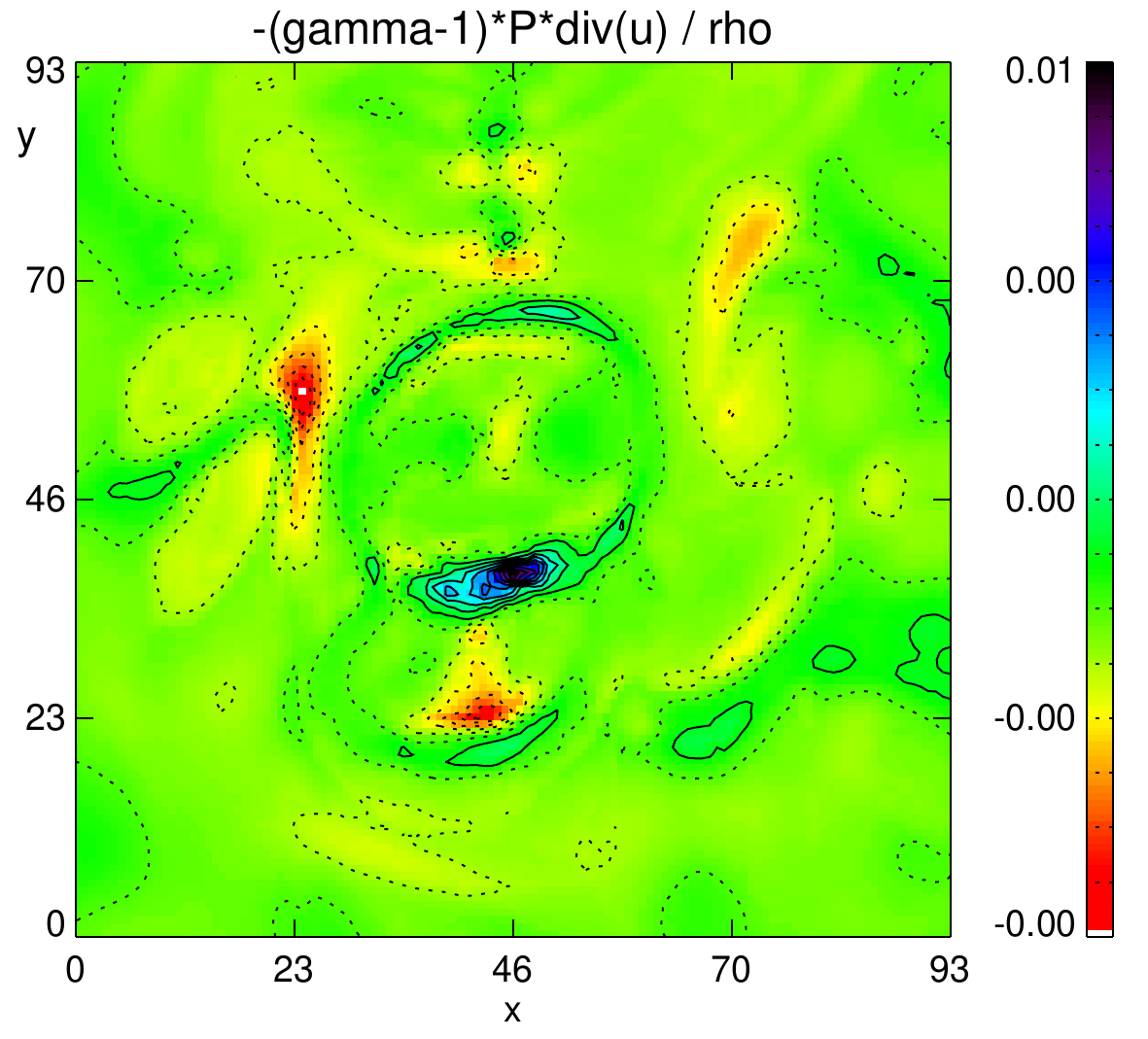}
\includegraphics[width= 0.25 \paperwidth]{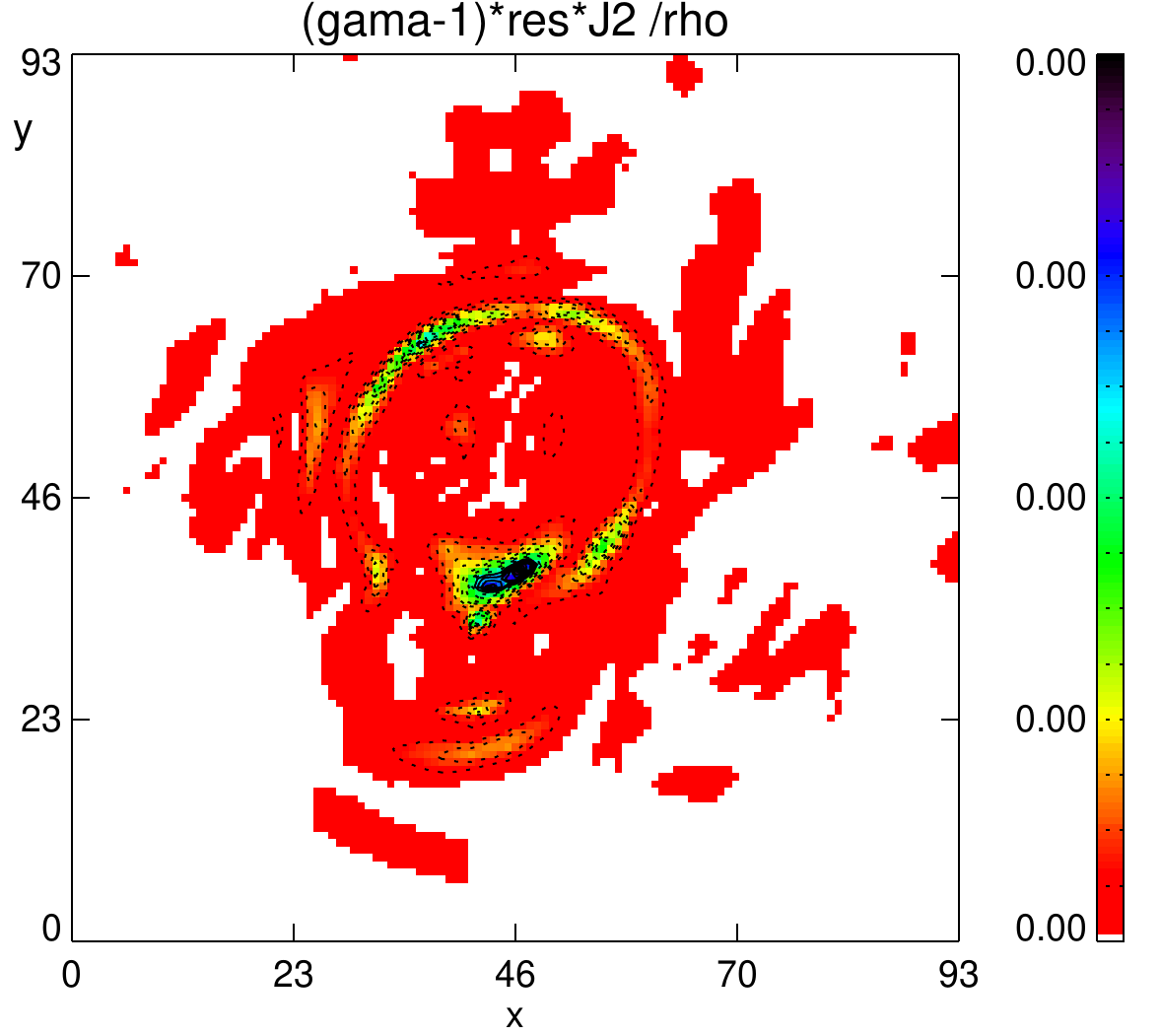}
\includegraphics[width= 0.25 \paperwidth]{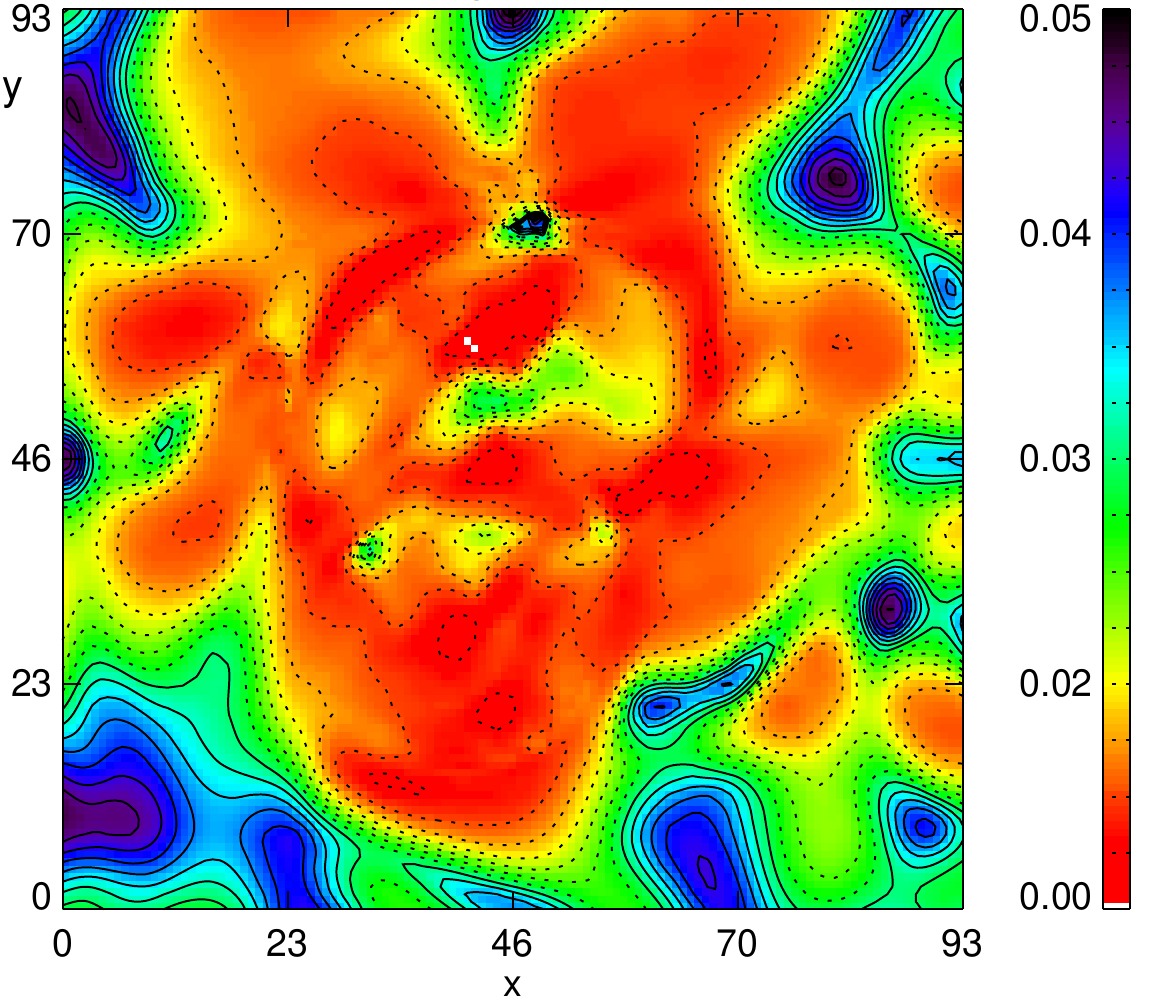}
\includegraphics[width= 0.25 \paperwidth]{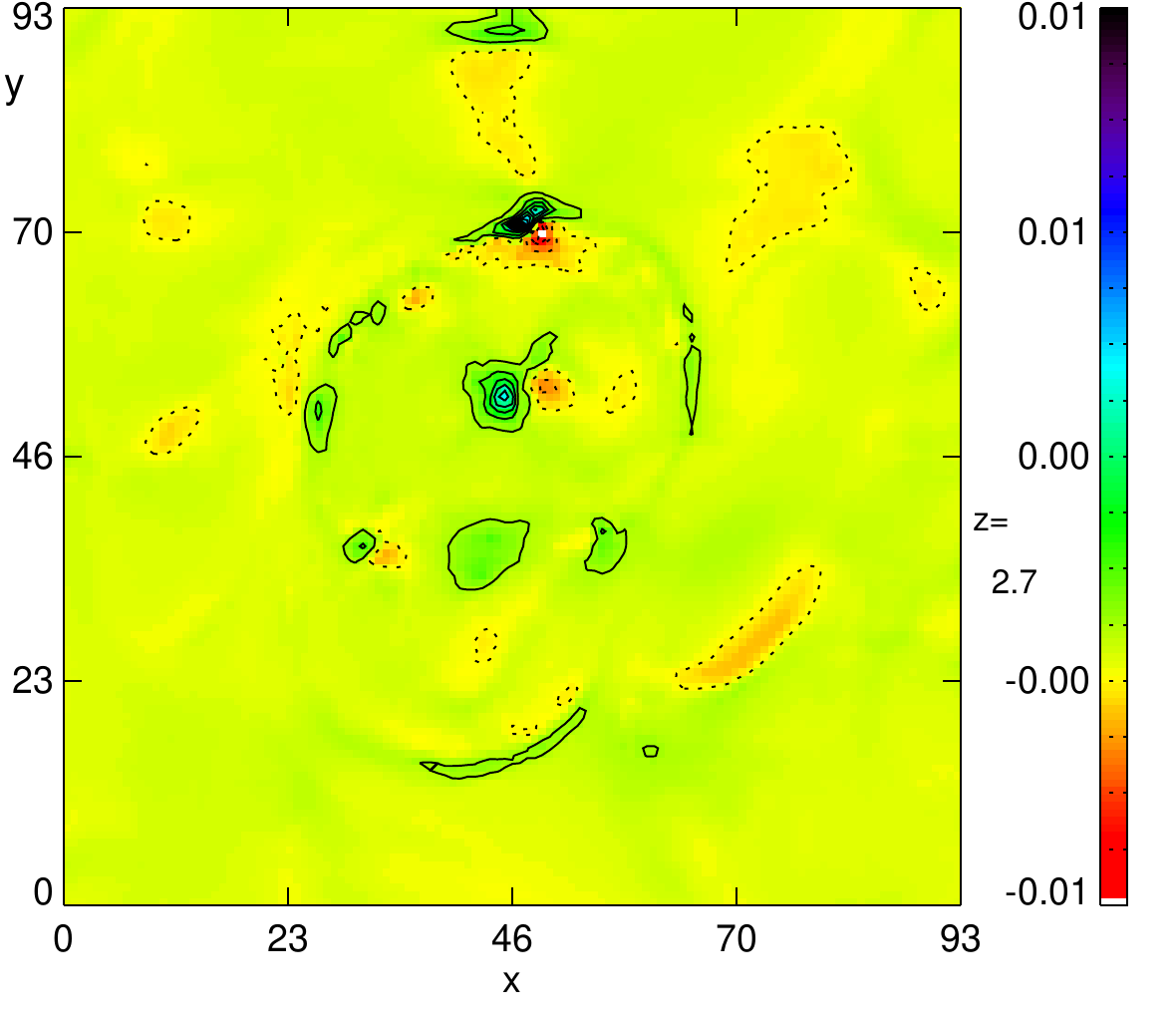}
\includegraphics[width= 0.25 \paperwidth]{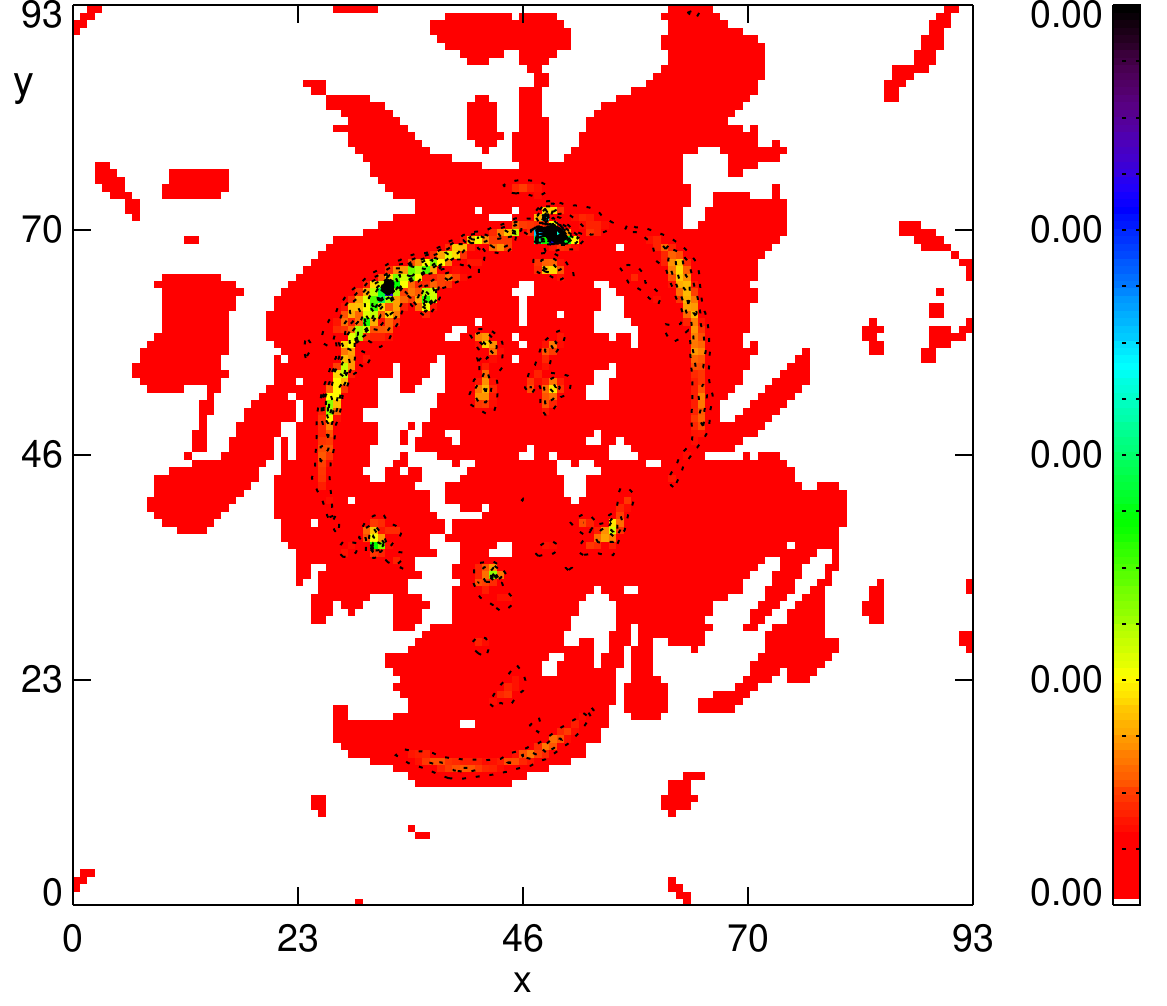}
   \caption{Temperature, first and second terms of Eq. 12 over density, in the left,
   middle and right panels, respectively. The result is shown at a horizontal plane in
   transition region at two instance of time, t = 80s in top and t = 160s in the bottom panels.    }
   \label{Tpj}%
\end{figure*}

The values of second term in the right hand side of the Eq. 11,
$(\gamma - 1) \ \eta j^{2} / \ \rho$, is shown in Fig.~\ref{gamresj2816} in
the plane x = 45.7 at t = 80s and t = 160s. By comparison with the
compressional part the contribution of the Joule heating appears to
be negligible. For a better comparison the contribution of the two
terms in the right hand side of the Eq. 11 in the temperature evaluation,
the horizontal view is shown at the height of transition region in two
different instance of time, t = 80 s in left and t = 160 s in the right panel of Fig.~\ref{Tpj}.
%
%%  for referee: 0.2
In the following, we will analyze in some more detail
to what degree compression and Joule heating contribute to the evolution of the
temperature. For this sake and in order to study the role of the forces involved
in the energy conversion process, we performed a volume integration
of the time rates of change of kinetic, magnetic and thermal
energies in the simulation box above the chosen Bright Point region. Our approach is similar
to \citet{birn2009} when they used energy transport equations to analyze the properties of energy conversions associated
with a reconnection process. The contribution of different terms in the
energy transport process can be studied from the following equations:
%
%Energy conservation: \\
     \begin{eqnarray}
%\begin{center}
\frac{d \varepsilon_{kin}}{d t} = - \frac{1}{2} \int_{S_{V}} \rho u^{2} \vec u \cdot d\vec s + \int_{V} (- \vec u \cdot \nabla p + \vec u \cdot \vec j \times
       \vec B) d^{3}v\\
\frac{d \varepsilon_{mag}}{d t} = - \frac{1}{\mu_{0}} \int_{S_{V}}(-\vec u \vec B^{2} + (\vec u \cdot \vec B)\vec B -\eta \vec
j \times \vec B) \cdot d\vec s  &\\ \nonumber   + \int_{V} (- \vec u \cdot \vec j \times \vec B - \eta \vec j^{2}) d^{3}v
\\
\frac{d \varepsilon_{th}}{d t} = - \frac{\gamma}{\gamma -1} \int_{S_{V}} p \vec u \cdot d \vec s + \int_{V} ( \vec u \cdot \nabla p + \eta \vec j^{2} ) d^{3}v
     \end{eqnarray}

\begin{figure}[htp]
        \centering
        \includegraphics[width= 8 cm]{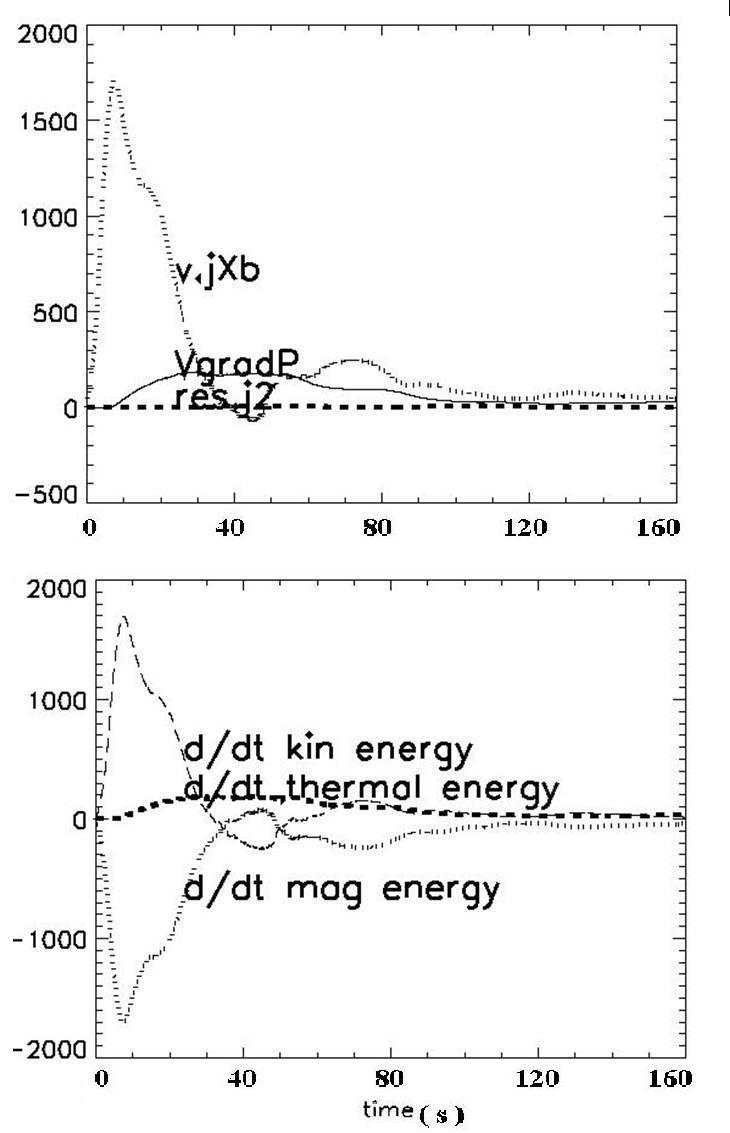}
        \caption{Rates of the contribution to the total energy changes, (bottom panel).
        Different contributions to the changes of the energy, (top panel).   }
        %The work done by Lorentz force, pressure gradient force and
        %Ohmic dissipation is shown in left panel, which are responsible for
        %the changes in the magnetic, thermal and kinetic energy rates (right panel).
                   \label{ddt1}%
         \end{figure}

Where $\varepsilon_{kin}$, $\varepsilon_{mag}$ and
$\varepsilon_{th}$ denote kinetic, $\rho u^{2} / 2$ , magnetic,
$\vec B^{2}/ 2 \mu_{0}$, and thermal, $ P / (\gamma -1)$, energies,
respectively. The volume integrals (second term on the right-hand
side) in these equations represent the energy conversion from one
form into the another. This energy conversion are explicitly written
in terms of the work done by Lorentz force, pressure gradient force
and Joule dissipation, (left panel of Fig.~\ref{ddt1}). Note
that the initial spike in the Lorentz force is in part caused by
numerical discretization errors and in part by the onset of
photospheric footpoint motion. The initial oscillations are damped
substantially during, approximately, two Alfv\'en times, followed by
a state of an approximate force balance. This effect was found to be
smaller in a run where footpoint motion was not included. The
initial perturbation has a minor effect on the initial extrapolated
magnetic field, it does not affect the currents and Lorentz forces
at a later times.

\begin{figure}[htp]
   \centering
\includegraphics[width= 7.8 cm, height = 0.2 \paperheight]{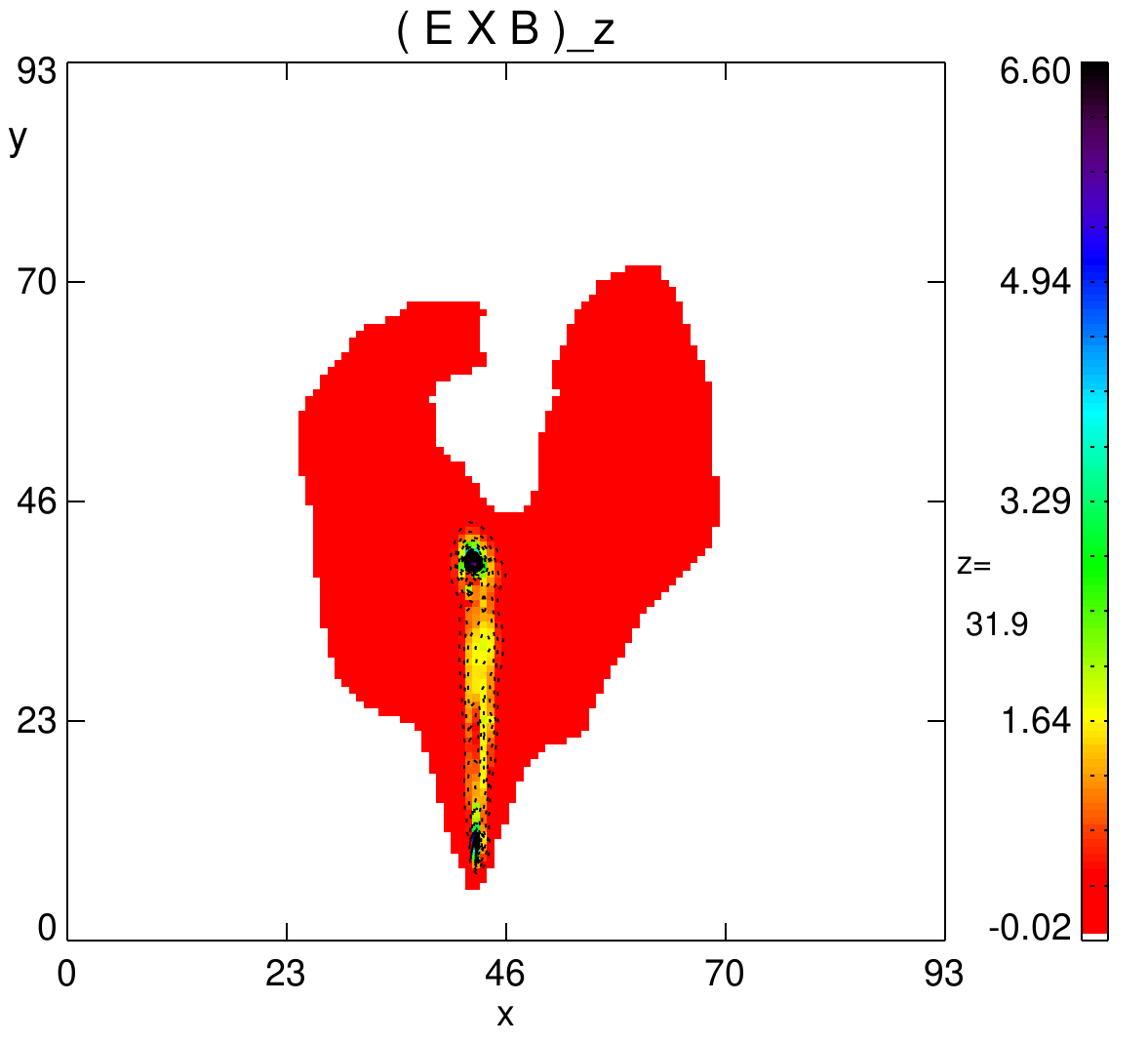}
\includegraphics[width= 7.8 cm, height = 0.2 \paperheight]{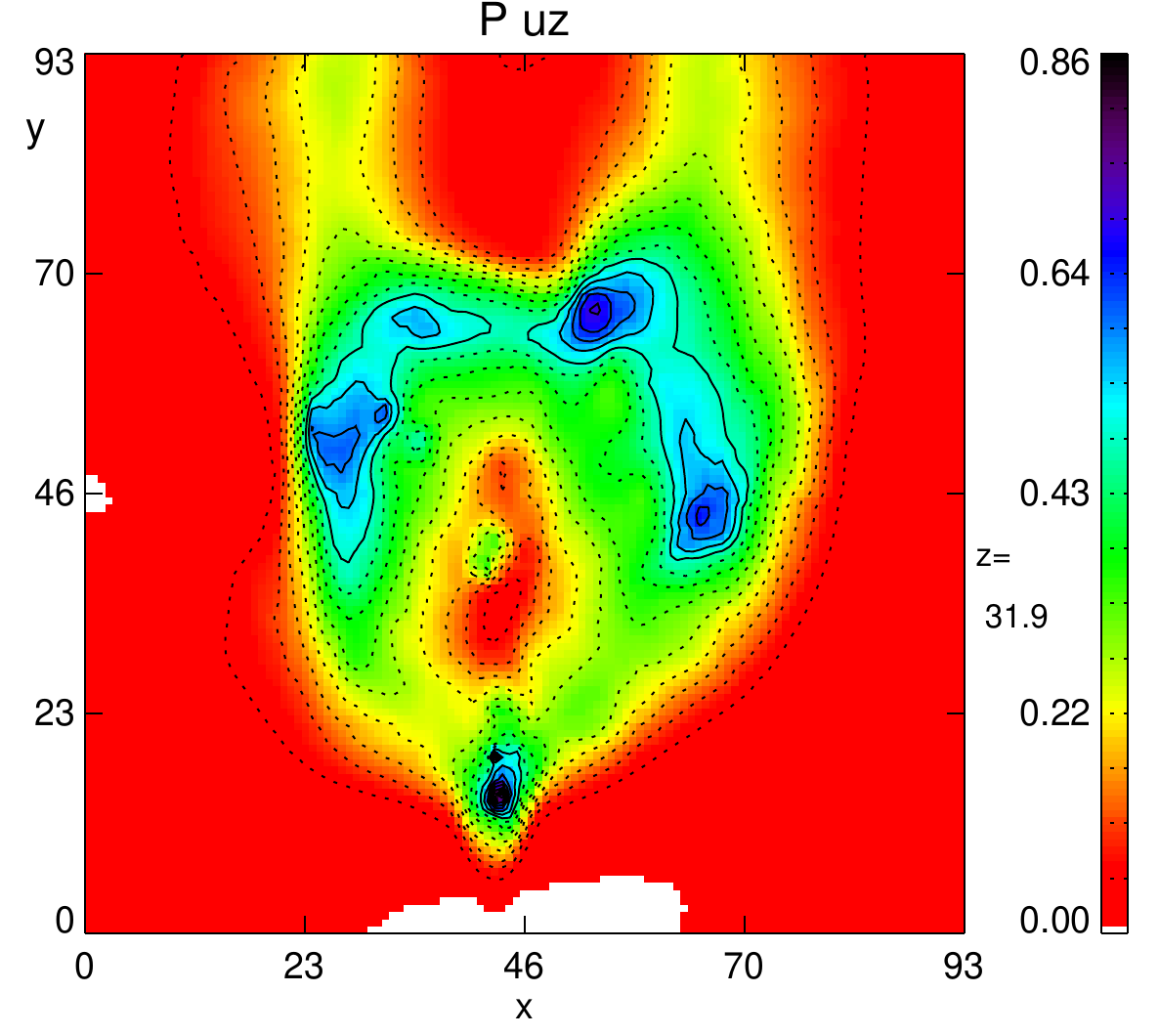}
\includegraphics[width= 7.8 cm, height = 0.2 \paperheight]{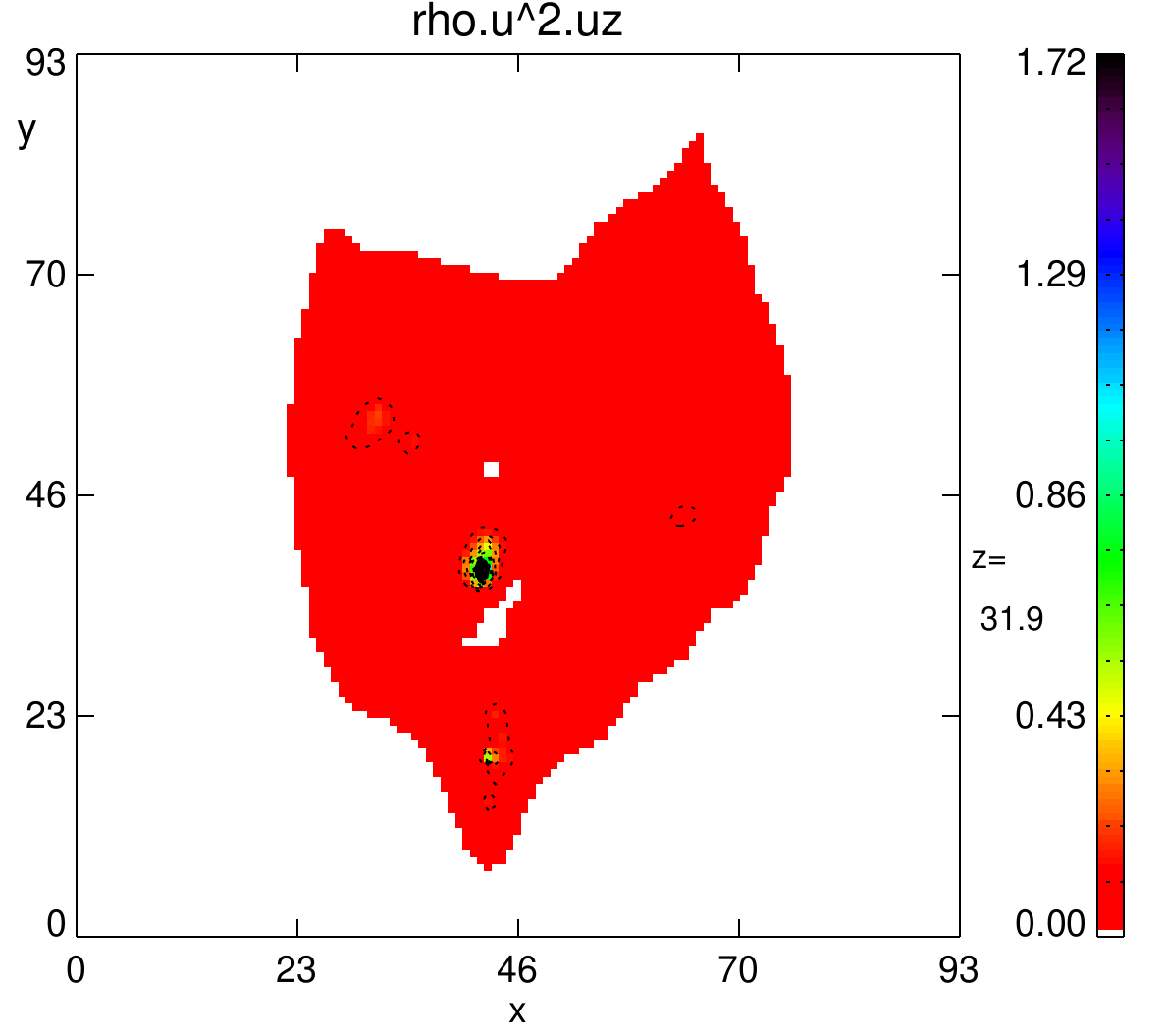}
   \caption{ First term in the right hand side of the Eqs.12-14, (surface integrals) is shown in
   the upper boundary of the simulation box after 160 s.         }
   \label{up}%
\end{figure}

The surface integrals are also needed
to obtain the energy rates, when they indicate the transport of
each of the three form of energies. With the chosen boundary condition however,
the values of these surface integrals are zero at the lower boundary. They
compensate each other through the side boundaries of the simulation
box as well. At the upper boundary however, one needs to consider the contribution of
this surface integrals in the rate of energy transfer. This means $\textbf{E} \times \textbf{B}$,
$P \textbf{u}$ and $\rho u^{2} \textbf{u}$ for the transport of magnetic, kinetic and thermal
energies, respectively. The values of these terms at the upper boundary are
shown in Fig.~\ref{up}. One can see that the contribution due to these terms is
insignificant, so it would be a good approximation to consider only the
volume integrals for the change in the energy rates.

%  fig: forces_eta_const.jpg ?

The changes in energy rates are shown in the right panel of
Fig.~\ref{ddt1}, the forces responsible for these changes are
depicted in the left panel of the Figure. As one can see by
comparing the two panels the magnetic energy is transferred to
kinetic energy almost completely via the work done by the Lorentz
force that accelerates the plasma. It is an intermediate step
however, followed by the work done by pressure gradient force that
converts kinetic energy into thermal energy. This decelerates the
plasma motion until, finally, the Lorentz force is balanced. The
direct transformation of magnetic energy to thermal energy (Joule
heating) is via Ohmic current dissipation, $ \eta J^{2}$. A
comparison of the energy conversions rates (see Fig.~\ref{ddt1},
right panel) however shows that Joule dissipation plays a minor role
in the energy exchange process while the other contributions are
orders of magnitudes larger. The minor role of Joule heating in
comparison to adiabatic process in the increase of thermal energy
was also found for the case of a solar flare by \citet{birn2009},
where they explained the compressional heating in two almost
simultaneously steps: acceleration by Lorentz force and deceleration
by pressure gradients.

\subsection{Influence of different resistivity models}
\label{resistivity} The previous calculation was based on an
anomalous resistivity model with the current carrier velocity as a
critical value for a local switch-on of additional resistivity. In
order to better understand the influence of the resistivity we
performed the simulation also with two other resistivity models, one
that uses a current density dependent resistivity and another with
constant resistivity respectively.

\begin{figure*}[htp]
     \centering
     \includegraphics[width= 0.8 \paperwidth]{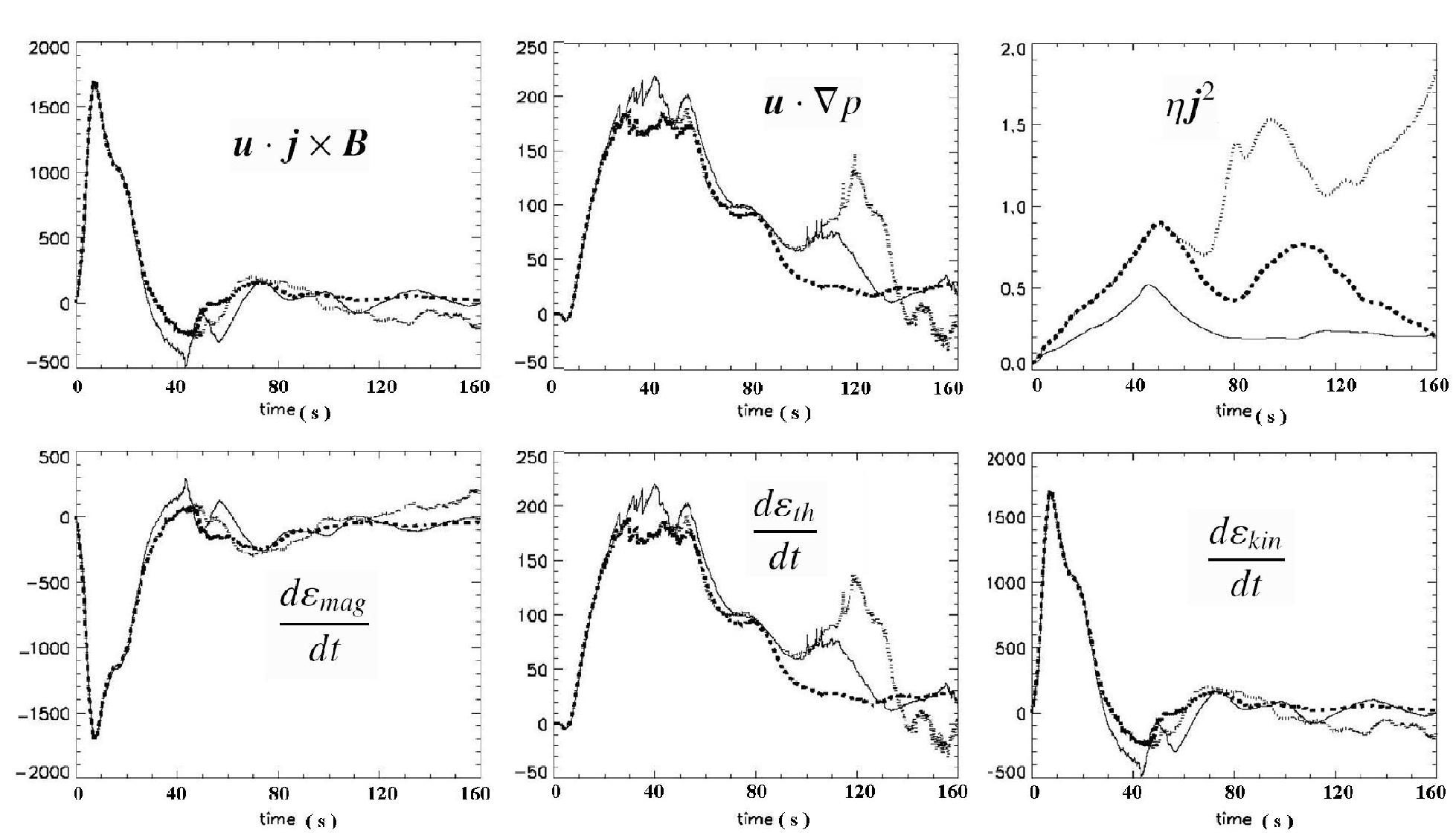}
    \caption{Energy change rates for three different resistivity models, (top panel).
    The bottom panel shows the work done by the Lorentz force, pressure gradient force and the Joule heating power.
    different lines correspond to anomalous current carrier dependent (dashed), anomalous current dependent (dotted)
    and constant (solid line) resistivity models.}
    \label{ddt}%
\end{figure*}

Fig.~\ref{ddt} depicts the resulting energy conversion rates and the
work done by the involved forces \ v $ \cdot \ J\times B$, v $ \cdot
\ \nabla P$ and by $\eta \ J^{2}$, for all the three resistivity models
by using different line styles for the results obtained by using the
different resistivity model. The results obtained for the three cases show that the resistivity
model influences the dynamics of the system and the thermal energy
rate mainly through the pressure gradient force. While magnetic and
kinetic energy rates of change depend only weakly on the resistivity
model, the rate of temperature change is significantly influenced.
Nevertheless, independent on the used resistivity model the heating
is due mainly to the work done by the pressure gradient force. At
the same time the contribution of the Joule heating is about two orders
of magnitude smaller (Note the scale of the plots in the top row.) We conclude that the adiabatic compression is
the dominant effect in increasing temperature in the BP region in
all three cases.
%! to include tot energies also use fig all.jpg

\subsection{Flux tube heating}
\label{fluxtube}

In order to locate the heating effect better it is appropriate to determine
it for individual flux tubes, integrating along the magnetic field lines
instead of taking values averaged over the whole simulation box as reported
in the previous sections. In this integration one has to take into account
the changing cross-section of flux tubes. This can be done by applying the
concept of the differential flux tube volume $V = \int{B}^{-1} ds$, where ds
indicates the step size along the field line. This way the flux conservation
in a flux tube ($\Phi = A \times B = const.$) is taken into account by the
proportionally of the cross-section to $B^{-1} $. Note that large flux tube
volumes correspond to field line rising high into the corona or hitting
regions of vanishing magnetic field. The energy is transported in accordance
with the upward directed Poynting flux $ E \times B$, enhanced magnetic
tension is carried away by wave propagation.

%   \includegraphics[width= 5 cm, height= 5 cm]{ujb}
%   \includegraphics[width= 5 cm, height= 5 cm]{ugp}
%  \includegraphics[width= 4.5 cm, height= 5 cm]{resj2_12}
%% a   \includegraphics[width= 5 cm, height= 5 cm]{resj2_12}
%   \includegraphics[width= 4.8 cm, height= 5 cm]{ddt_mag_12}
%% a   \includegraphics[width= 4.8 cm, height= 5 cm]{ddt_kin12}
%   \includegraphics[width= 5 cm, height= 5 cm]{ddt_kin12}
%   \includegraphics[width= 5 cm, height= 5 cm]{ddt_th12}
%%  for referee 0.2
\begin{figure*}[htp]
   \centering
   \includegraphics[width= 0.25 \paperwidth,height= 4 cm ]{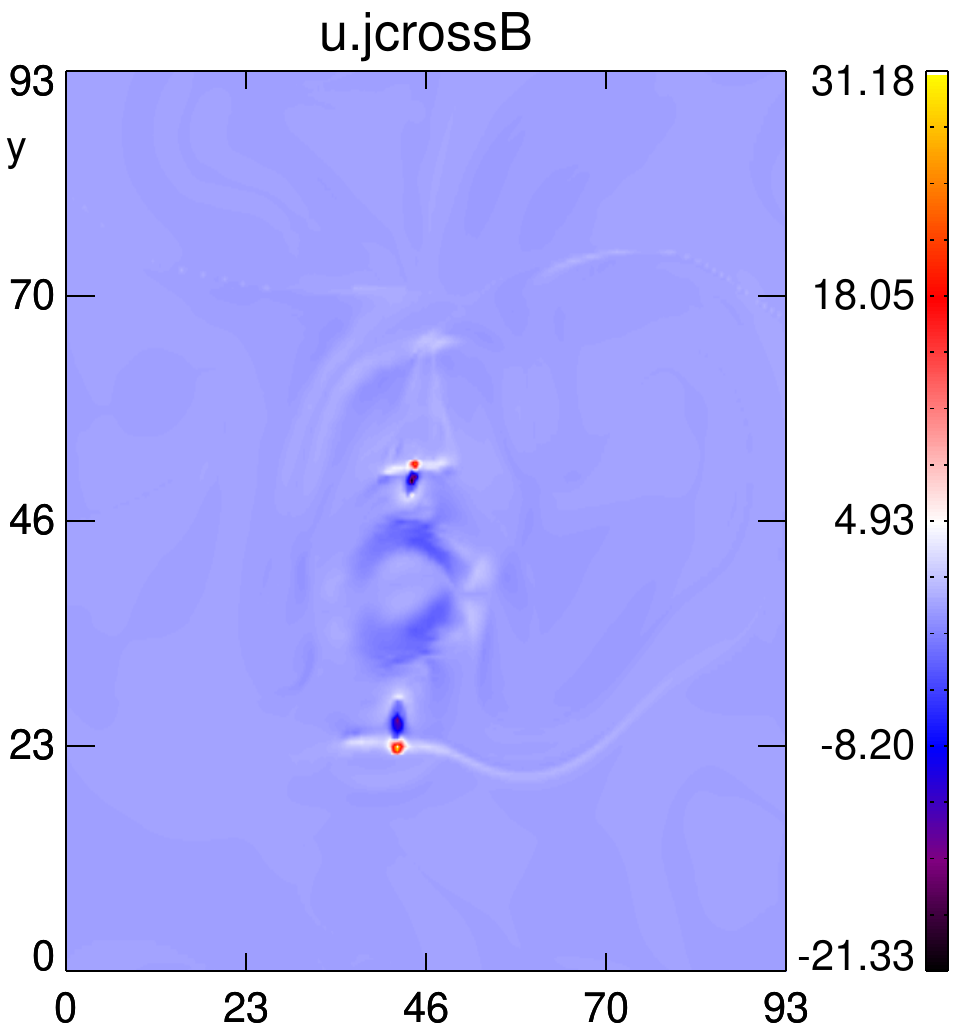}
   \includegraphics[width= 0.25 \paperwidth, height= 4 cm]{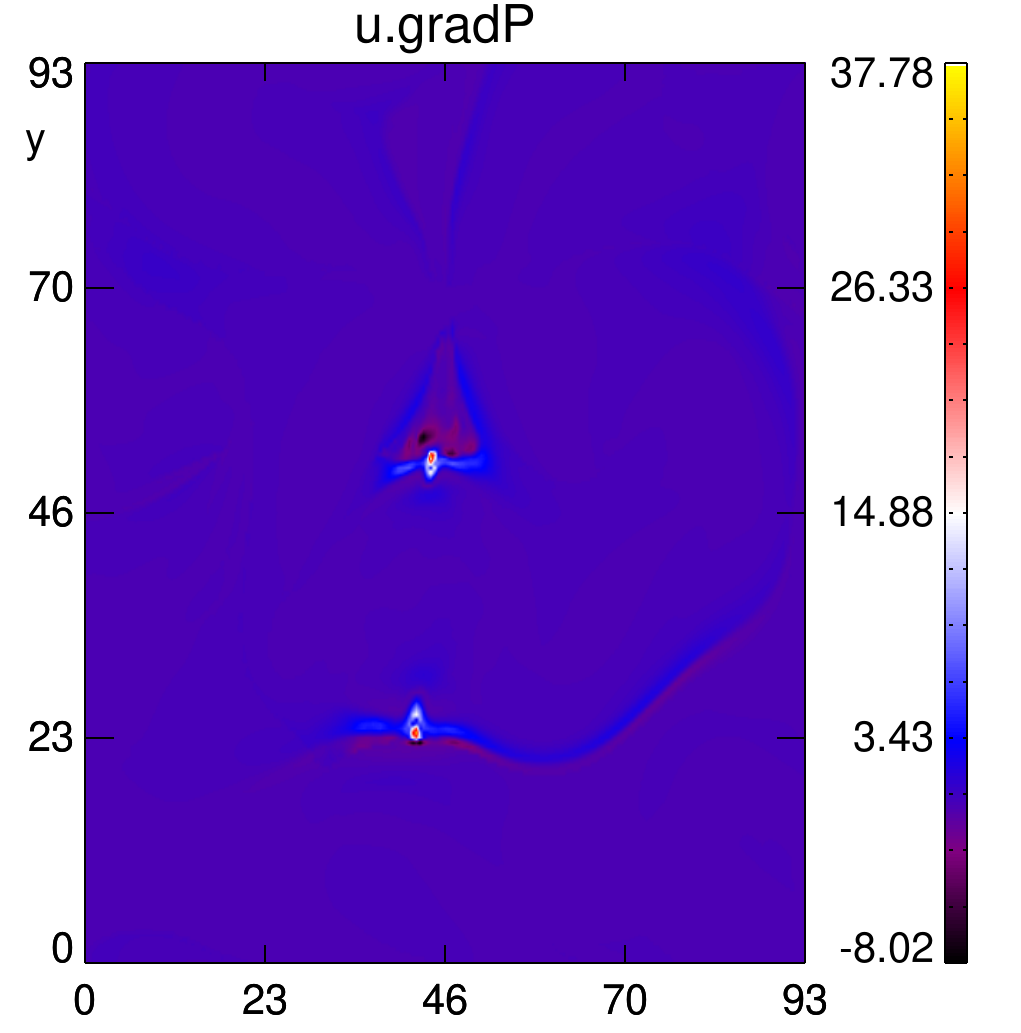}
  \includegraphics[width= 0.25 \paperwidth, height= 4 cm]{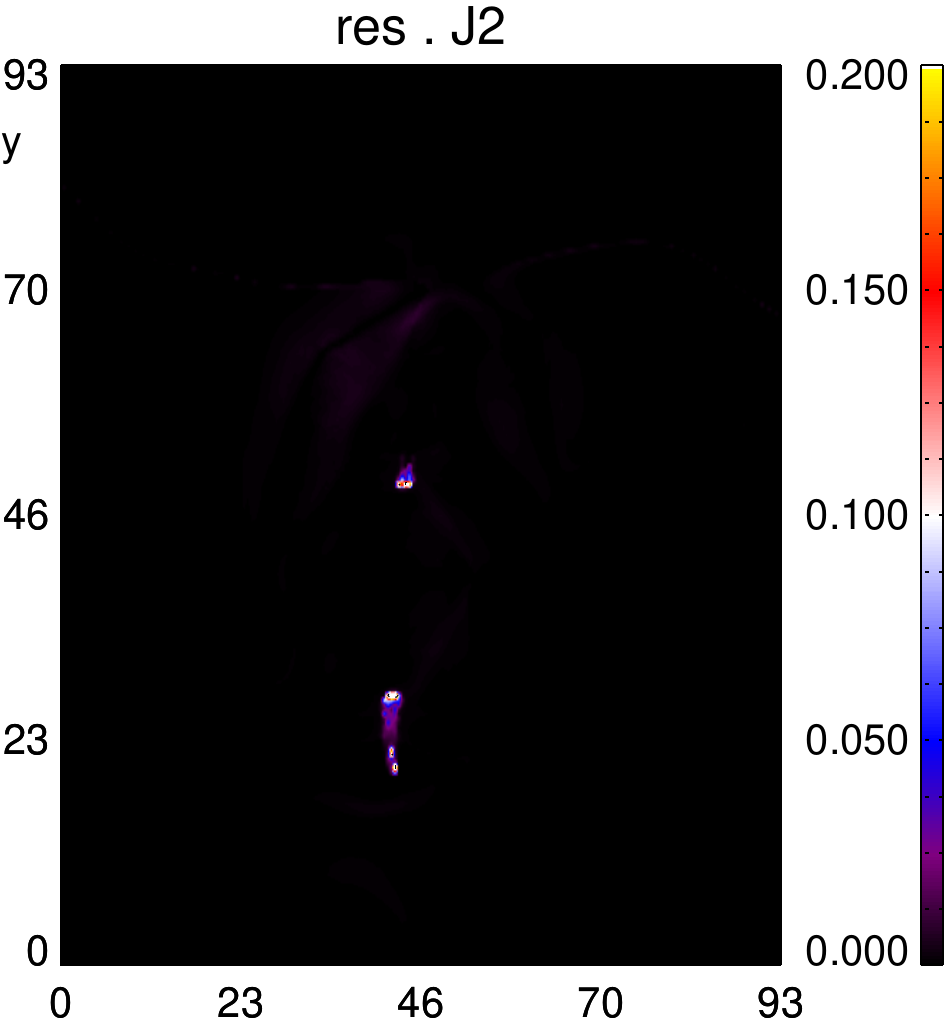}
   \includegraphics[width= 0.25 \paperwidth, height= 4 cm]{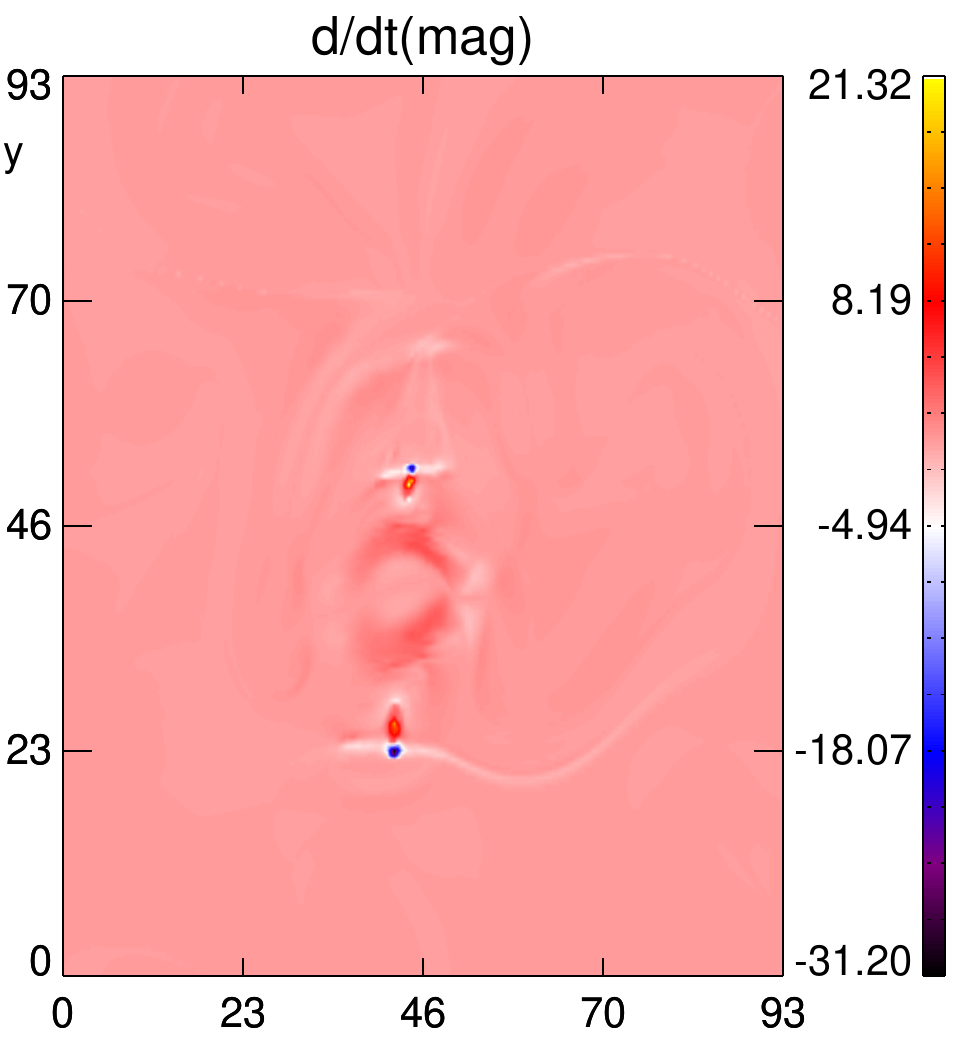}
   \includegraphics[width= 0.25 \paperwidth, height= 4 cm]{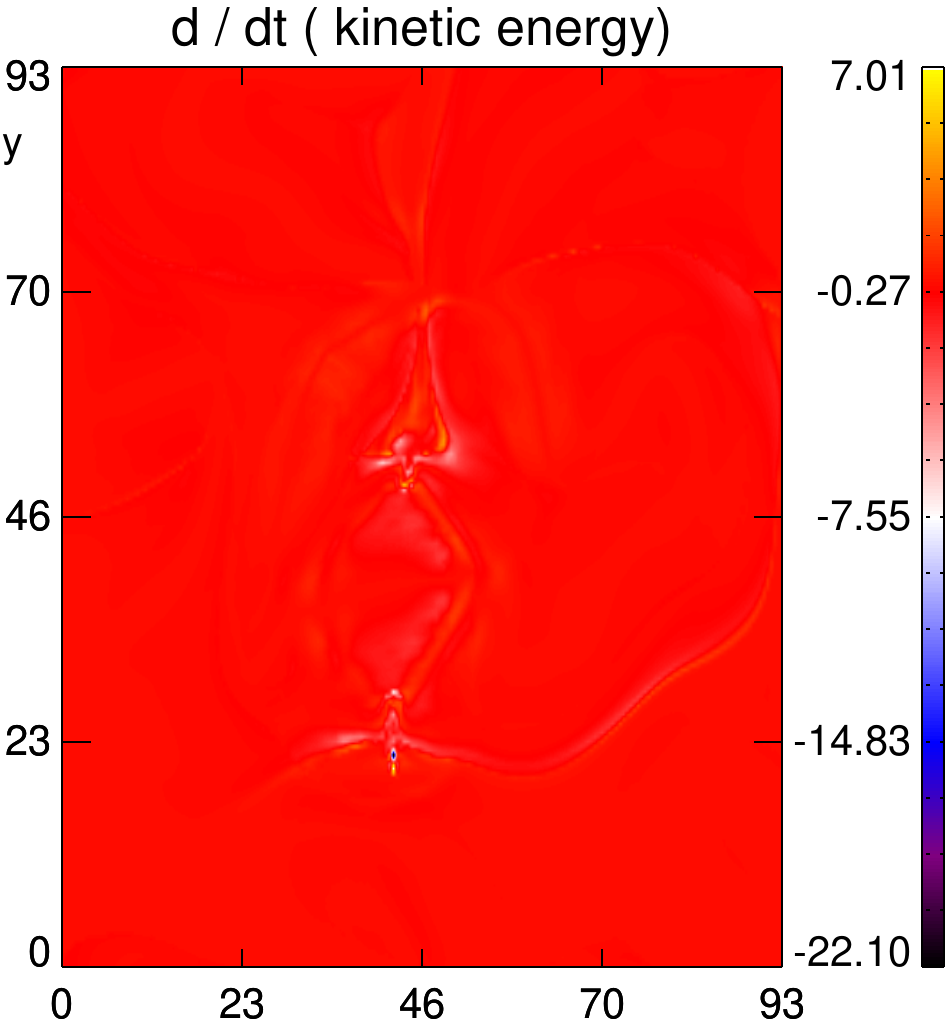}
   \includegraphics[width= 0.25 \paperwidth, height= 4 cm]{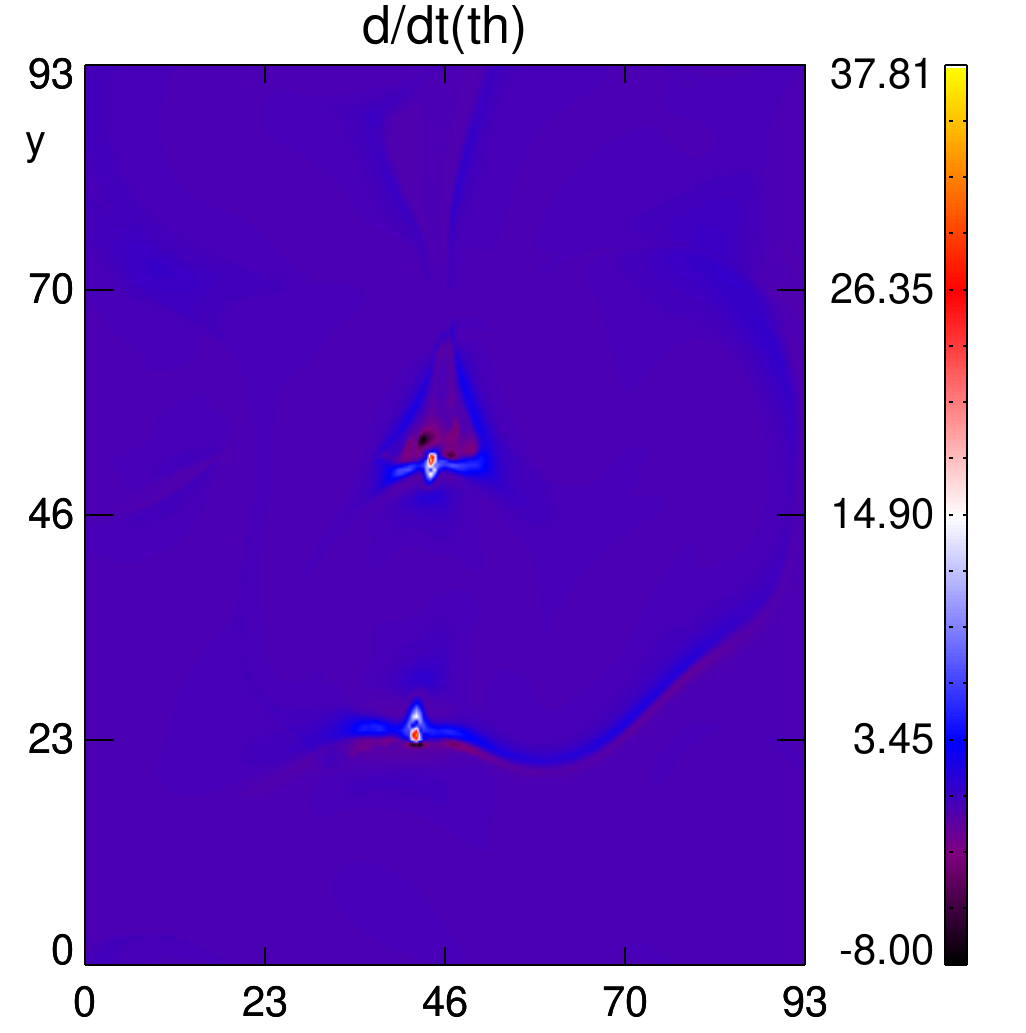}
   \caption{ Integration along the field lines using the differential flux
tube volume concept for the work done by Lorentz force, pressure
gradient force and Joule heating, (top panel, from left to right),
and the changes in rates of magnetic field change, the temperature
and the kinetic energy, (bottom panel, from left to right) at t =
160s. }
              \label{ujbugpdt}%
    \end{figure*}

\begin{figure}[htp]
   \centering
   \includegraphics[width= 0.2 \paperwidth]{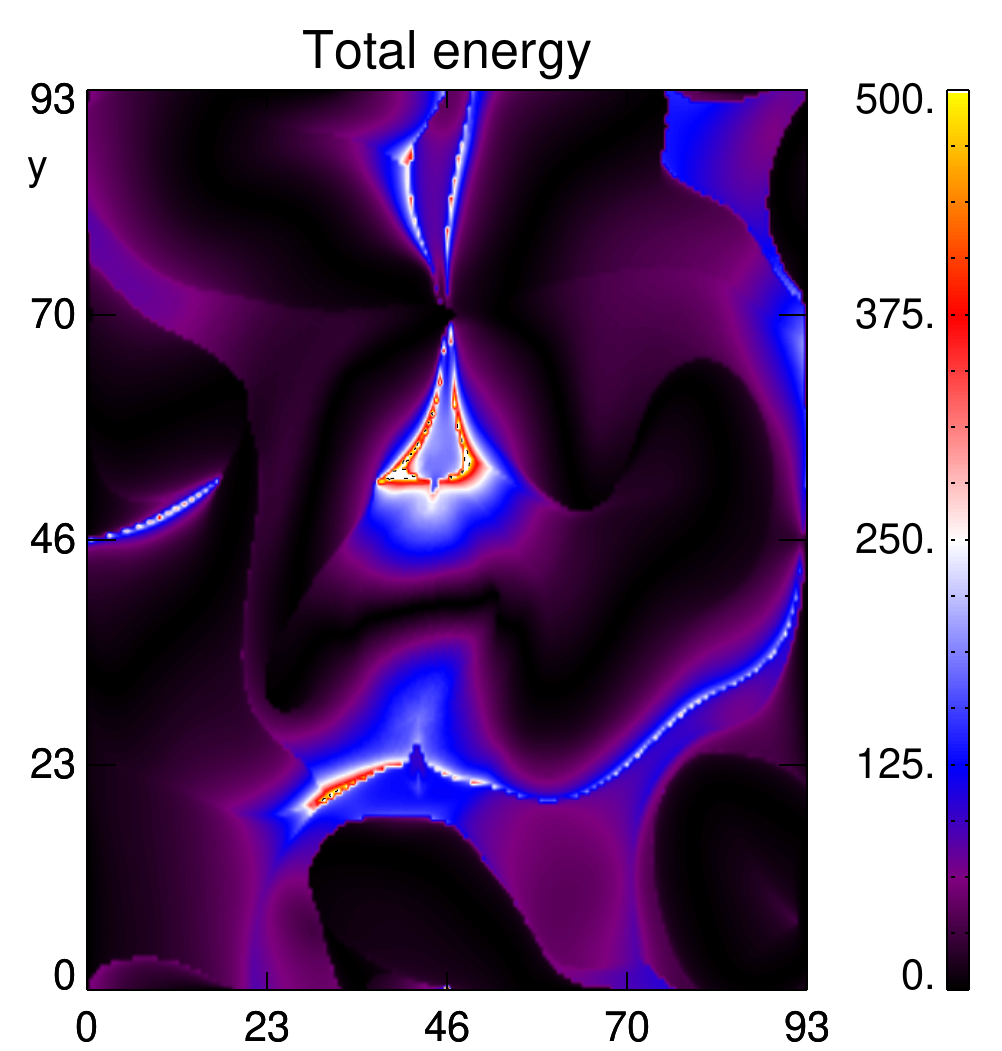}
   \includegraphics[width= 0.2 \paperwidth]{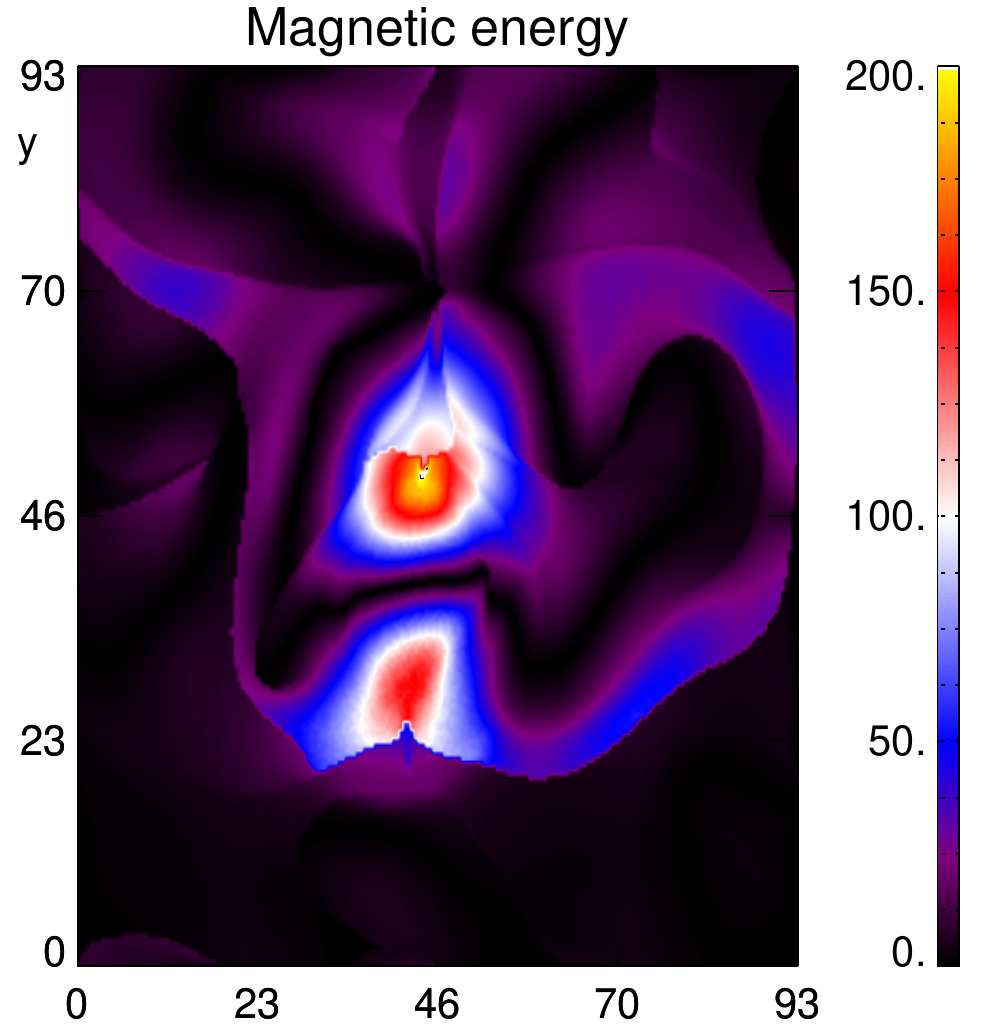}
   \includegraphics[width= 0.2 \paperwidth]{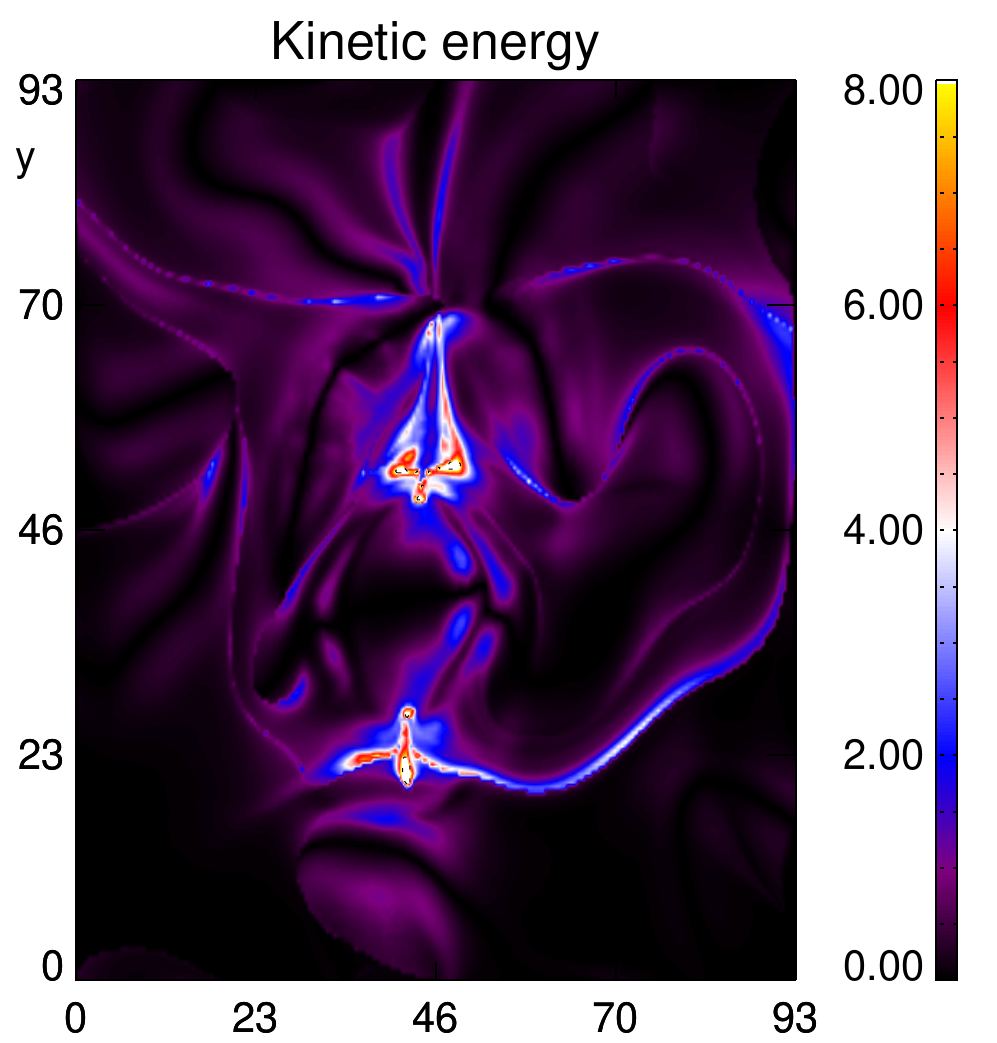}
   \includegraphics[width= 0.2 \paperwidth]{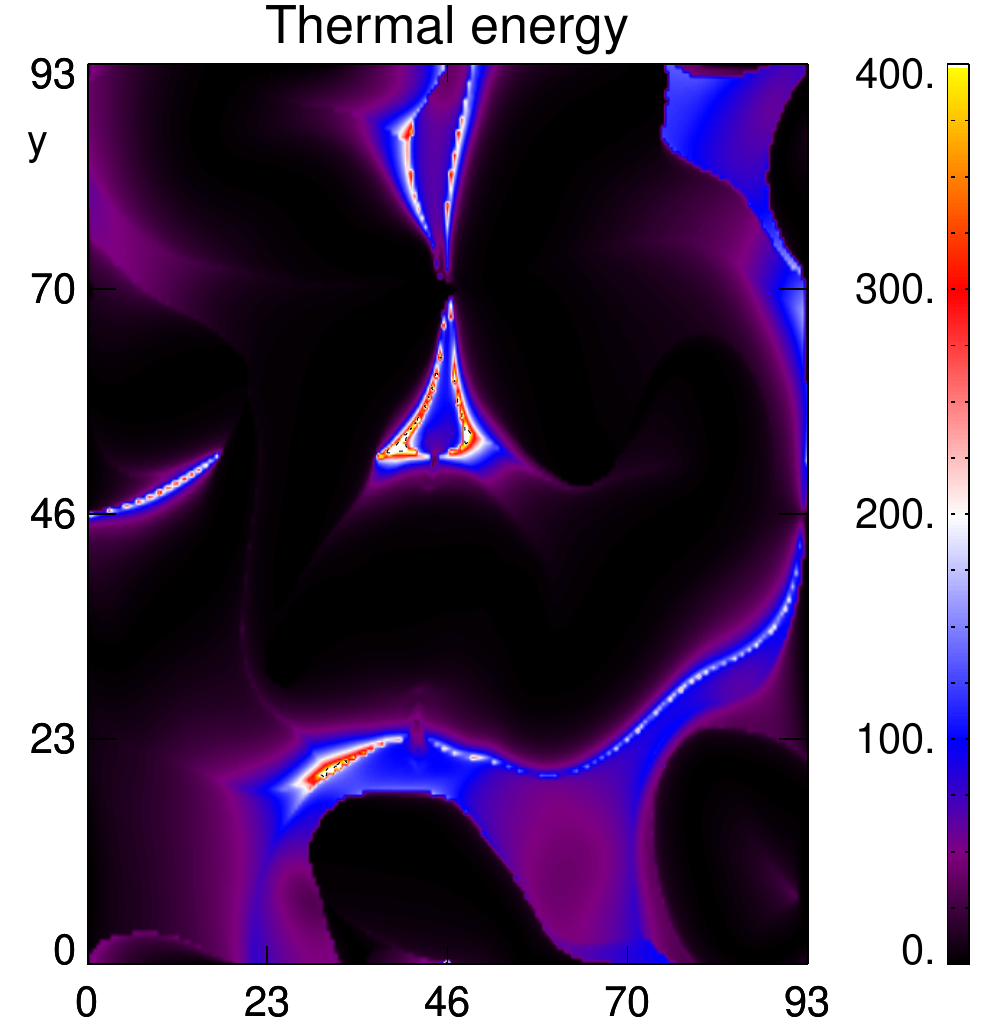}
   \caption{ Results of integration along the magnetic field lines using the differential flux
tube volume concept for total energy(top, left panel) and magnetic(top, right panel), kinetic(bottom,
left panel) and thermal energies(bottom, right panel), at t = 160s.     }
              \label{tot}%
    \end{figure}

\begin{figure}[htp]
   \centering
 \includegraphics[width= 0.2 \paperwidth]{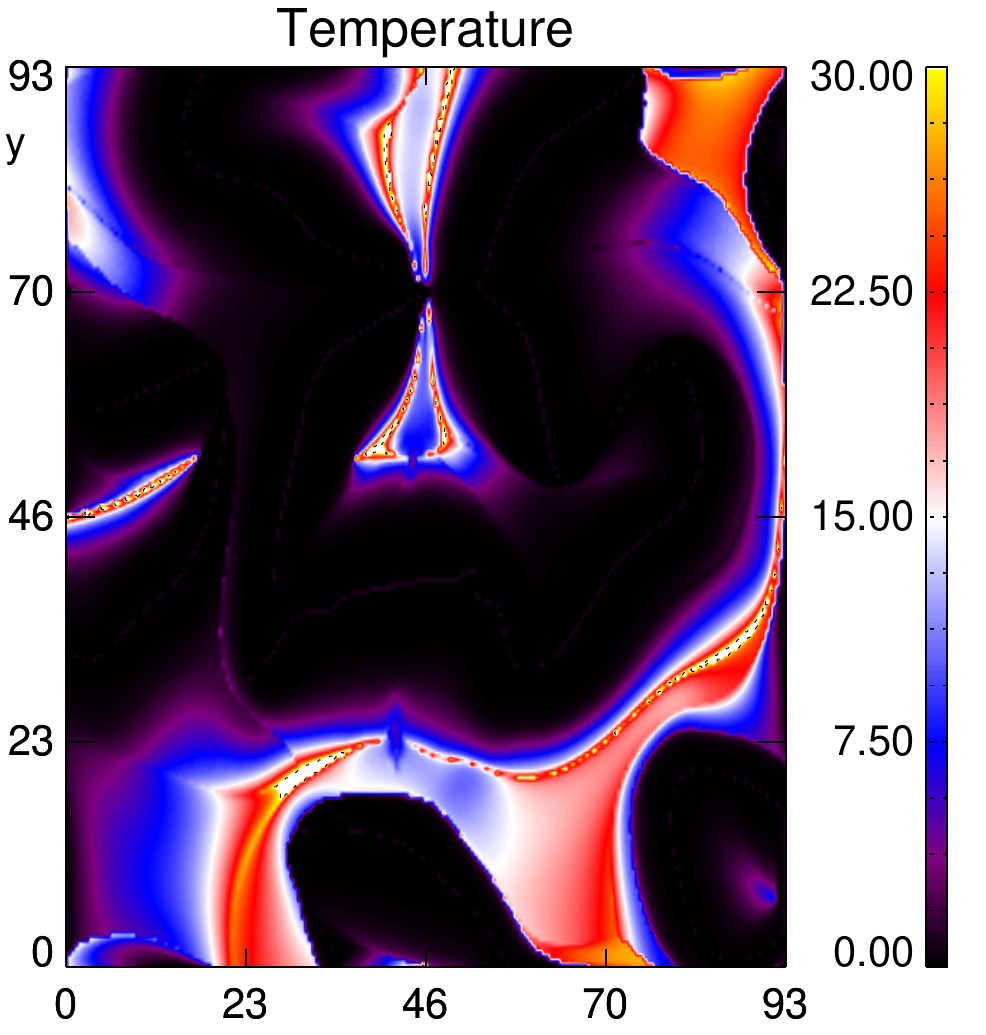}
 \includegraphics[width= 0.2 \paperwidth]{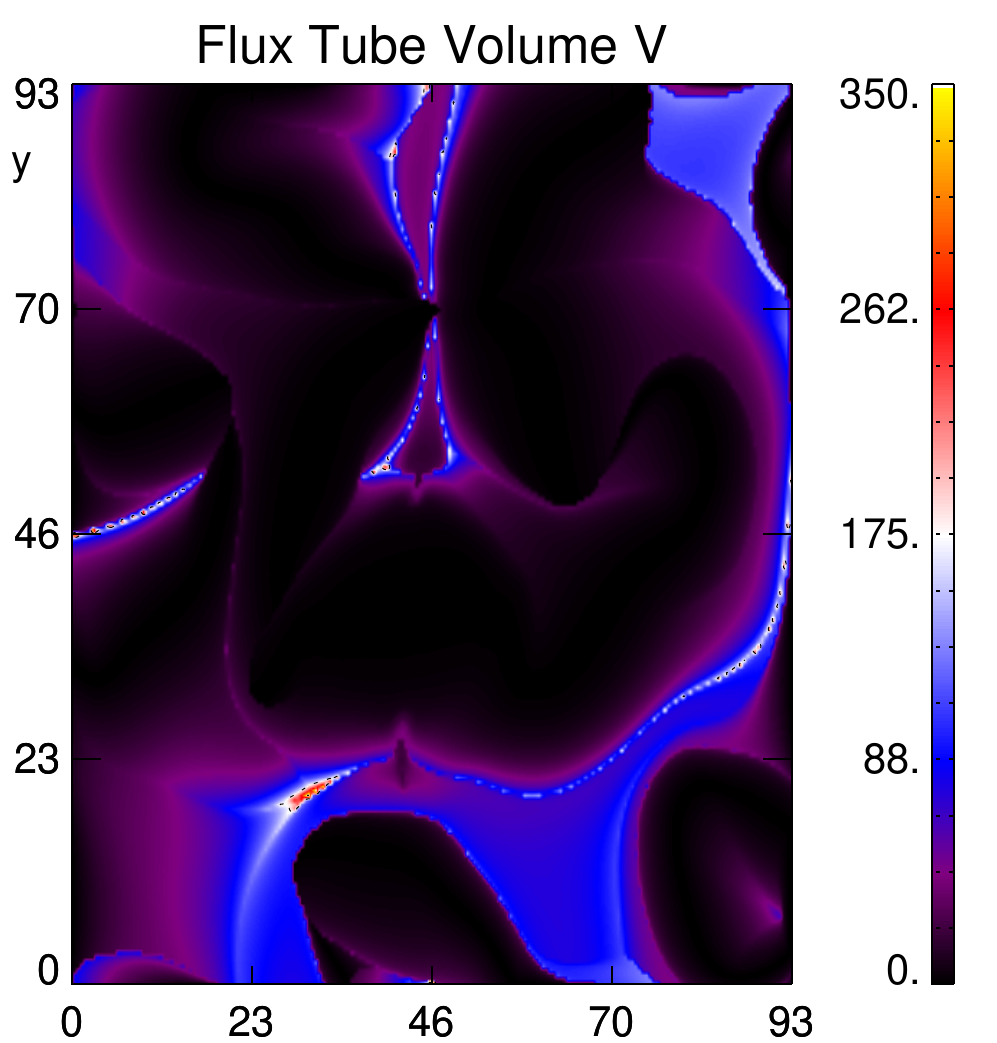}
   \caption{ Temperature and flux tube volume, integrated along the magnetic
field lines, at t = 160s.   }
              \label{Tvol}%
    \end{figure}

For the quantities described in section~\ref{global} the resulting flux tube
integrated values are shown in Fig. ~\ref{ujbugpdt} in the horizontal
reference plane just above the transition region. The values reached indicate
once more the negligible role of Joule heating by current dissipation for the
thermal energy change in the bright point region compared to the dominant
role of the pressure gradient force. Please note the different range of the
plots in Figure~12 as indicated by the color bar. It also can be seen that
the locations at which this force and also maximum rates of energy changes
appear coincide. Furthermore, the same pattern has formed in the integration
result of v $ \cdot \ \nabla P $, v $ \cdot \ J \times B $ and the rate of
change of the different kinds of energy. This pattern can clearly be seen in
the integration of total energy along the field lines (Fig. ~\ref{tot}),
which is the sum of the kinetic, magnetic and thermal energies:
$$ \varepsilon = \varepsilon_{kin} + \varepsilon_{mag} + \varepsilon_{th}
= \int_{V} \{ \frac{1}{2} \rho u^{2} + \frac{1}{2 \mu_{0}} B^{2} + \frac{p}{\gamma -1} \} d^{3}v
$$
%%  for referee: 0.25
%\vspace{5 mm}
The left panel of Fig.~\ref{Tvol} shows the result of this
integration for temperature and flux tube volume. The coincidence of
the temperature enhancement with the maxima obtained in the flux
tube integrated energy change rates and forces shows that the heat
is provided by the plasma compression due to the Lorentz force.

Enhanced flux tube integrated values follow the same pattern as the
BP. This indicates that the regions of enhanced temperatures
correspond to the foot points of field lines leading to higher
altitudes or to regions where the magnetic field vanishes. The
plasma motion across these regions supplies the magnetic energy that
is converted to thermal energy.

%----------------------------------
\section{Summary and discussion}
\label{summary}

We have presented the results of heating processes in the region of
an observed X-ray coronal bright point. In particular we have
investigated the importance of the work done by adiabatic
compression in comparison with Joule heating in the course of the
dynamic evolution and heat production near the bright point.

The simulation shows that an arc-shaped structure of enhanced
temperature forms that is 2-4 times hotter than the background
plasma. This structure is located above the two main opposite
photospheric magnetic flux concentration. It coincides with the
location where the electrical current densities are maximum. The
structures of temperature and current density enhancements, indeed,
coincide.

We further examined the contribution of the Lorentz force, pressure
gradient force and Joule heating performing volume integrals in the
simulation box that determine the magnetic, kinetic and thermal
energy change rates for three different resistivity models. We found
that independent on the resistivity model magnetic energy was
transformed to kinetic energy through the work done by Lorentz
force. Kinetic energy in turn is converted to thermal energy due to
pressure gradients that balance the Lorentz force.

A comparison of the effect of the three energy conversion through
v  $ \cdot \ J\times B$, v  $ \cdot \ \nabla P$ and $ \eta \ J^{2}$ show that adiabatic
compression has an important role in temperature increase in the upper corona.
This is not dependent on the resistivity model used in the simulation.

For a better understanding of the heating processes we utilized
the concept of differential flux tube integration of the different
contributions along the magnetic field lines. A quantitative
comparison in the horizontal plane, from where the integration
starts, shows that energy conversion rate, total energies and work
done by Lorentz and pressure gradient forces are located in the
same flux tubes, also temperature and flux tube volume are maximum
at the same place.

We conclude that the conversion of magnetic energy to kinetic energy
via the work done by the Lorentz force and from kinetic to thermal
energy due to the work done against the pressure gradient force
determine the heating of this bright point. We could show that
plasma compression dominates the heating of the bright point. In
contrast, the role of Joule dissipation appeared to be negligibly
small. The temperature enhancement follows the same pattern. The
fact that the pattern obtained by calculating flux volume integrals
coincides with the one of temperature and energy change rates bring
us to the conclusion that plasma motion at the footpoints of the
flux tubes carries the energy upward and makes the flux tubes rise
to the higher corona. The magnetic energy is converted to thermal
energy until the plasma compression is balanced by the Lorentz
force. In the local, flux-tube oriented consideration we also could
see that the role of the Joule heating in these energy conversion
processes was negligible and the heating of plasma in the bright
point region is basically due to pressure gradient force.

\textbf{First, the fact that Joule heating is weak in the corona was
not entirely unexpected but it is quantitatively confirmed here. It
is worth to remember that the necessary up-scaling of the
resistivity and of the onset condition of micro-turbulent anomalous
resistivity to the resolved by the MHD simulation grid scales does
even overestimate the actual Joule heating. As a result Joule
heating cannot be considered a viable process unless there is a
convincing argument that the dissipation regions are volume filling
to a much larger extend than the already large one used in the
present model.
%using models that already vastly overestimate the size of the reconnection (Joule heating) regions.
\\
Second, the results demonstrate very clearly that compression is an
important processes in the energy budget. It is not clear in how far
compression can contribute to the overall coronal heating but it is
certainly important for the local heating of BPs.
\\
Third, in this context the nature and the consequences of plasma
compression are worth some consideration. In ideal MHD adiabatic
compression is reversible. But the consequent flux tube heating is,
however, irreversible due to magnetic reconnection and other mixing
processes. Magnetic reconnection, in particular, changes flux tube
identities (magnetic connectivity) while flux tube entropy
conservation requires ideal MHD in addition to appropriate boundary
conditions. Local adiabatic compression becomes irreversible also
due to other plasma transport processes like  heat conduction and
radiative cooling. These aspects will be separately investigated in
a subsequent paper. Meanwhile the results presented here clearly
demonstrate that in the overall energy budget plasma compression
(and expansion) can play an important role in the heating of the
corona.}

\begin{acknowledgements}
One of the authors (S.J.) gratefully acknowledges her
Max-Planck-Society PHD-stipend.
\end{acknowledgements}

\end{document}